\documentclass[11pt]{article}
\usepackage{amsmath}
\usepackage{graphicx}
\usepackage{enumerate}
\usepackage{natbib}
\usepackage{url} 

\usepackage{dsfont}
\usepackage{bm}

\newcommand{\bg}{{\bm g}}
\newcommand{\bx}{{\bm x}}
\newcommand{\bxs}{{\bm x}^*}
\newcommand{\bSigma}{{\bm \Sigma}}
\newcommand{\bS}{{\bm S}}
\newcommand{\btheta}{{\bm \theta}}
\newcommand{\bbeta}{{\bm \beta}}
\newcommand{\blambda}{{\bm \lambda}}
\newcommand{\mydot}{{\, . \,}}
\newcommand{\minus}{\scalebox{0.75}[1.0]{$-$}}

\title{Flexible models for nonstationary dependence: Methodology and examples}
\author{Benjamin D. Youngman\\ Department of Mathematics, University of Exeter, UK}

\begin{document}

\maketitle

\section{Introduction}
When modelling spatial processes, it may be inappropriate to assume that dependence is both stationary and isotropic, especially when dealing with large domains. For example, when modelling US rainfall, we might expect a different dependence structure over mountainous regions from over plains. This work allows nonstationarity in dependence by deforming the space on which a process is typically defined to one in which stationarity and isotropy are reasonable assumptions. Formally consider $\bx \in \mathds{R}^p$ and a mapping $\bg : \mathds{R}^p \mapsto \mathds{R}^q$ such that $\bxs = \bg(\bx)$ for $\bxs \in \mathds{R}^q$. \cite{sampson1992} introduced this approach in the context of spatial modelling, so that $p=q=2$: then $\bx \in \mathds{R}^2$, a coordinate in geographic space (henceforth $G$-space), is mapped to $\bxs \in \mathds{R}^2$, a coordinate in dispersion space (henceforth $D$-space), where $\bxs = \bg(\bx)$ for $\bg : \mathds{R}^2 \mapsto \mathds{R}^2$. Stationarity and isotropy are assumed for $D$-space. The remainder of this work focuses on $\bx \in \mathds{R}^2$. 

\cite{sampson1992} represent $\bg$ as a pair of thin plate splines, which are estimated by multidimensional scaling. Subsequent works by \cite{damian2001} and \cite{schmidt2003} adopt a Bayesian approach to inference and assume that $\bg$ is a random function and data are realisations of a Gaussian process (GP); i.e., \[Y_t(\bx) \mid \bg \sim GP\big(\mu(\bx), v\big(\bg (\bx), .\big)\big)\] for some process $Y_t(\bx)$ at time $t=1, \ldots, T$, location $\bx \in \mathds{R}^2$, and mean and covariance functions $\mu$ and $v$, respectively. For a fixed set of locations, $\bx_1, \ldots, \bx_n$, this allows inference to be performed through the likelihood \begin{equation}L({\bm \Sigma}) = |2 \pi \bSigma|^{-(T - 1) / 2} \exp\Big(\minus\dfrac{T}{2}\text{tr}(\bSigma^{-1} {\bm V})\Big),\label{gplik}\end{equation} where ${\bm V} = T^{-1}\sum_{t=1}^T({\bf y}_t - \hat {\bm \mu})({\bf y}_t - \hat {\bm \mu})^T$, ${\bf y} = ({\bf y}_1, \ldots, {\bf y}_T)$ with ${\bf y}_t = (y_t(\bx_1), \ldots, y_t(\bx_n))$, $\bSigma$ has $(i, j)$th element $\Sigma_{i, j} = v(\bg (\bx_i), \bg (\bx_j))$ and $\hat {\bm \mu} = (\hat \mu_1, \ldots, \hat \mu_n)$ where $\hat \mu_i = T^{-1}\sum_{t=1}^T y_t(\bx_i)$. \cite{damian2001} and \cite{schmidt2003} place thin plate spline and Gaussian process priors on $\bg$, respectively, and use Markov chain Monte Carlo to sample from the posterior distribution of $\bg$, which readily allows its uncertainty to be quantified. 

Spatial deformation models suffer the intuitively undesirable flaw of allowing $D$-spaces that `fold', i.e., non-bijective mappings $\bg$ or, more conceptually, mappings such that for every $\bx$ in $G$-space there is not a single $\bx^*$ in $D$-space. \cite{damian2001} hinder folding by considering the ``bending energy'' of $D$-spaces, and increasingly penalize spaces that require less energy to bend, or equivalently are more prone to deviate from the affine transformation. Alternatively, \cite{schmidt2003} propose to represent $\bg$ as a multivariate Gaussian process, and state that ``the GP formulation for [$\bg$] tends to eliminate the kind of non-injective mappings that were noted by \cite{sampson1992}''. 

Various approaches have explicitly addressed avoiding mappings that fold. \cite{iovleff2004}, for example, use a Delaunay triangulation of $G$-space locations to identify and eliminate mappings that give rise to folds in $D$-space. \cite{perrin-mon} derive conditions on deformations based on radial basis functions that avoid folds. Nonstationarity in dependence is also considered when emulating computer models and referred to as input warping (IW): the computer model's inputs are transformed to a scale on which dependence is stationary; see, e.g., \cite{snelson2004}. In recent work, \cite{zammit} propose deep compositional spatial models for representing $\bg$ in which the compositional formulation can ensure bijectivity. \cite{zammit} propose to represent the compositions through IW GPs and deep stochastic processes (DSPs), both of which are based on basis representations with weights and unknown basis function parameters. The IW GPs have unknown weights, which are estimated by maximum likelihood, whereas the DSPs have random weights, which are assumed to be of log-Gaussian form and estimated by variational Bayes. \cite{zammit} then propose three approaches to warping: axial warping units, with positive weights and monotonic basis functions; radial basis functions, employing the constraints of \cite{perrin-mon}; and M\"{o}bius transformation units, which make analogy between mapping from $\mathds{C}$ to itself with mapping from $\mathds{R}^2$ to itself. Each can be used with IW GPs and DSPs and ensure bijectivity by virtue of the compositional structure. 

Folding could be considered a consequence of a two-dimensional $D$-space being insufficient to bring isotropy. \cite{schmidt2011} and \cite{bornn2012} propose extending $D$-space to $2 + r$ dimensions for $r \geq 1$ so that $\bg : \mathds{R}^{2} \mapsto \mathds{R}^{2 + r}$. \cite{bornn2012} refer to this approach as dimension expansion. \cite{schmidt2011} place a GP prior on $\bg$, allow covariates in $v$ and then base $v$ on Mahalanobis distance, which generalizes the usual Euclidean distance. \cite{bornn2012} estimate the latent dimensions in a two-stage procedure that finds interim values using a least squares fit between empirical and model-based variograms, which are then approximated using thin plate splines.

In this work, the next section introduces flexible models for nonstationary dependence based on the spatial deformation and dimension expansion approaches. Section 3 introduces objective methods of inference for such models. Section 4 demonstrates the proposed modelling framework on the solar radiation data originally used in \cite{sampson1992}. Section 5 presents a case study on risk due to extreme rainfall, in which extreme rainfall over part of Colorado, US, as studied in \cite{cooley2007}, is modelled and then simulated. Section 6 summarizes the work presented.

\section{Methodology} \label{meth}

Consider again $Y_t(\bx)$, values of some phenomenon at time $t=1, \ldots, T$ and location $\bx \in \mathcal{G}$. The two-dimensional case of $\bx = (x_1, x_2)$, where $x_1$ and $x_2$ are longitude and latitude coordinates, respectively, shall be considered. This readily extends to $G$-spaces defined over any number of dimensions, as in \cite{bornn2012}. Spaces also need not be defined geographically: see \cite{cooley2007} for the notion of `climate space'. Independence over time will be assumed to focus on spatial dependence.

\subsection{General framework} \label{spatial}

Spatial processes with a dependence structure fully characterized by a dependence function will be considered, which can be denoted by \begin{equation} \label{spatial} Y_t(\bx) \mid \bg \sim SpatialProcess\big(v(\bg(\bx), \mydot)\big).\end{equation} A zero-mean Gaussian process, as studied in \cite{damian2001}, and also presented in \S\ref{S:solar} and \S\ref{S:rain}, is one example of such a process. The notation $v(\bg(x), \, . \,)$ implicitly represents $v(\bg(x), \, . \,; {\bm \theta})$, for some dependence parameters $\bm \theta$. Suppression of $\bm \theta$ facilitates focusing on estimating $\bg$. Estimation of $\bm \theta$ is deferred to \S\ref{S:inf:pen}.

\subsection{Nonstationary covariance $v($\emph{g}$($\emph{x}$), \mydot)$} \label{S:meth:nonstat}

The two approaches of spatial deformation and dimension expansion will be considered for introducing nonstationarity into $v(\bx, \, . \,)$. The following synthesizes notation previously introduced for spatial deformation and dimension expansion models.

Spatial deformations and dimension expansions are both represented as $\bx^* =\bg(\bx)$ for mapping $\bg$. For spatial deformations $\bg:\mathds{R}^2 \mapsto \mathds{R}^2$ where $\bx^* = (x_1^*, x_2^*) = (g_1(\bx), g_2(\bx))$. For dimension expansions $\bg:\mathds{R}^2 \mapsto \mathds{R}^{2 + r}$ where ${\bm x}^* = (\bx / \phi, g_1(\bx), \ldots, g_r(\bx))$, for $\phi > 0$. In both cases $g_d : \mathds{R}^2 \mapsto \mathds{R}$ and $\bx$ and $\bx^*$ exist in $G$- and $D$-space, respectively. Both cases will also use a dependence function of the form $v(\bx,  \bx') = \gamma(||\bg(\bx) - \bg(\bx')||)$ for covariance function $\gamma$.

The spatial deformation and dimension expansion approaches each have pros and cons. A particularly attractive feature of the former is its interpretability: $D$-space can be visualized in two dimensions, which in turn may simplify relating regions of relatively long- or short-range spatial dependence to known phenomena. Such interpretation is less immediate for dimension expansions as each dimension must be visualized in three dimensions. How to intuitively represent three-or-more-dimensional spaces, or combinations of dimensions, e.g., of $z_1$ and $z_2$, is not immediate. Projections on to lower dimensions, as explored in \cite{schmidt2011} may be beneficial, but are not explored here. The dimension expansion approach naturally avoids non-bijective transformations, and could be seen to be more flexible by allowing $D$-space to be of any dimension, unlike in spatial deformations where $D$-space is limited to $\mathds{R}^2$. Choosing between a spatial deformation or a dimension expansion is therefore a trade-off between interpretability and flexibility: the decision is likely to depend on the application.

\subsection{A finite-rank basis representation for \boldmath{$g$}} \label{S:meth:basis}

Finite-rank spline-based forms for $\bg$ are chosen here, such that \begin{equation} g_d(\bx) = \sum_{k=1}^{K_d}\beta_{dk} b_{dk}(\bx),\label{basis} \end{equation} where $\beta_{dk}$ are basis coefficients and $b_{dk}$ are basis functions. This linear form in $\bbeta_d = (\beta_{d0}, \ldots, \beta_{dK_d})'$ means that $x_1^*$ and $x_2^*$ in the spatial deformation model, or $z_d$ in the dimension expansion model, can be written as ${\bf x}_* {\bm \beta}_*$, where ${\bf x}_*$ corresponds to a row of a design matrix ${\bf X}_*$ with elements determined by the $b_{*k}$ basis functions. As this form also applies to $\bx / \phi$ in the dimension expansion model, estimating $\phi$ can be absorbed into estimating $\bg$. 

\cite{sampson1992}, \cite{schmidt2003} and \cite{bornn2012} have previously used thin plate splines to define the $b_{dk}$ basis functions. Here it is proposed to use \emph{regression} splines, with a focus on thin plate regression splines \citep{wood-tprs}. These are based on representing the $g_d$ that would be obtained from thin plate splines, i.e., with knots at each location, through eigenbases obtained from a truncated eigendecomposition. This gives an optimal finite-rank representation of $g_d$ relative to its full-rank counterpart, and better performance for rank $K_d$ than a thin plate spline with $K_d$ knots. Here the thin plate regression splines are extended for deformations to incorporate the extra constraints derived in \cite{smith1996}, which avoid rotationally invariant deformations. Instead of thin plate regression splines, other two-dimensional basis functions could be used for the $b_{kd}$s, or they could be formed through tensor products of lower-dimensional splines \citep{deboor1978, wood-tensor}. For example, a two-dimensional basis can be formed from the tensor product of two one-dimensional bases. This flexibility allows deformations to be characterized similarly to smooths in generalized additive models (GAM); see, e.g., \cite{wood-book}.

\section{Inference} \label{S:inf}

Estimation of the spatial deformation or dimension expansion models will be presented for a fixed set of locations, $\mathcal{X} = \{\bx_1, \ldots, \bx_n\}$, and a fixed set of time points, $\mathcal{T} = \{1, \ldots, T\}$. Corresponding data are ${\bf y} = ({\bf y}_1, \ldots, {\bf y}_T)$, where ${\bf y}_t = (y_t(\bx_1), \ldots, y_t(\bx_n))$. Fitting either model corresponds to estimating the dependence parameters, $\bm \theta$, basis coefficients, $\bm \beta$, which determine the $G$- to $D$-space mapping $\bg$, and some smoothing parameters, $\bm \lambda$. As mentioned in \S\ref{meth}, $\btheta$ can be absorbed in $\bbeta$ so that the spatial process model of relation \eqref{spatial} has a log-likelihood $\ell(\bbeta)$. 

\subsection{Roughness penalized likelihood} \label{S:inf:pen}

Various spline-based representations for $g_d$ lead naturally to roughness penalties of the form $\bbeta_d^T \bS_d \bbeta_d$, which penalize wigglier $g_d$ more, where $\bS_d$ is a penalty matrix, with elements determined by the $b_{dk}$ basis functions. A smoothing parameter $\lambda_d > 0$ typically multiplies the roughness penalty to control the amount of smoothing. Concatenating the $\bbeta_d$s into the vector $\bm \beta$, and the $\lambda_d \bS_d$s into a block diagonal matrix $\bS_\blambda$, gives a penalized log-likelihood of the form \begin{equation*} \label{penalized} \ell_{p_0}(\bbeta, \blambda) = \ell(\bbeta) - \frac{1}{2}\bbeta^T \bS_\blambda \bbeta,\end{equation*} which allows estimation of $\bbeta$ given smoothing parameters $\blambda$. If $\btheta$ is absorbed in $\bbeta$, for example into $\bbeta_d$, then $\bS_d$ is supplemented with rows and columns of zeros corresponding to where $\btheta$ is in $\bbeta_d$; see \cite{wood-reml}. In previous works smoothing parameters have been considered as bending energies: larger values lead to surfaces that need more energy to bend.

If thin plate splines are used with a null space comprising linear terms in $x_1$ and $x_2$ (see, e.g., \cite{wood-tprs}), then $\lambda_d \to \infty$ leads to an affine transformation. For dimension expansion, it may be preferred that $\lambda_d \to \infty$ corresponds to $z_d({\bm x}) \to 0$ for all $\bx$. This can be achieved by modifying the penalty matrix, $\bS_d$, according to its zero eigenvalues: see \cite{marra2011}.

\subsection{Penalising folding (for deformations only)} \label{S:inf:fold}

A drawback to spatial deformations, raised in \S\ref{S:meth:nonstat}, is that they may ``fold'', i.e., produce non-bijective mappings between $G$- and $D$-space such that a point in $G$-space maps to multiple points in $D$-space. This is likely to be unintuitive for some situations and may want to be avoided. \cite{bornn2012} demonstrate how smoothing parameters associated with thin plate splines, or equivalently that control the bending energy, may be fixed to ensure bijectivity. This work aims to maintain objectivity by allowing optimal estimation of smoothing parameters while ensuring bijectivity. \cite{iovleff2004} ensure bijectivity by representing $G$-space as a Delaunay triangulation, which, when transformed to $D$-space, is bijective if none of the vertices lie within any of the triangles. The approach of \cite{iovleff2004} applies to any form for $\bg$, which is a criterion that the approach proposed here also satisfies. The approaches of \cite{perrin-mon} and \cite{zammit} require specific---albeit seemingly rather flexible---forms for $\bg$.

\begin{figure}[h!]
\begin{center}
\includegraphics[width=.98\textwidth]{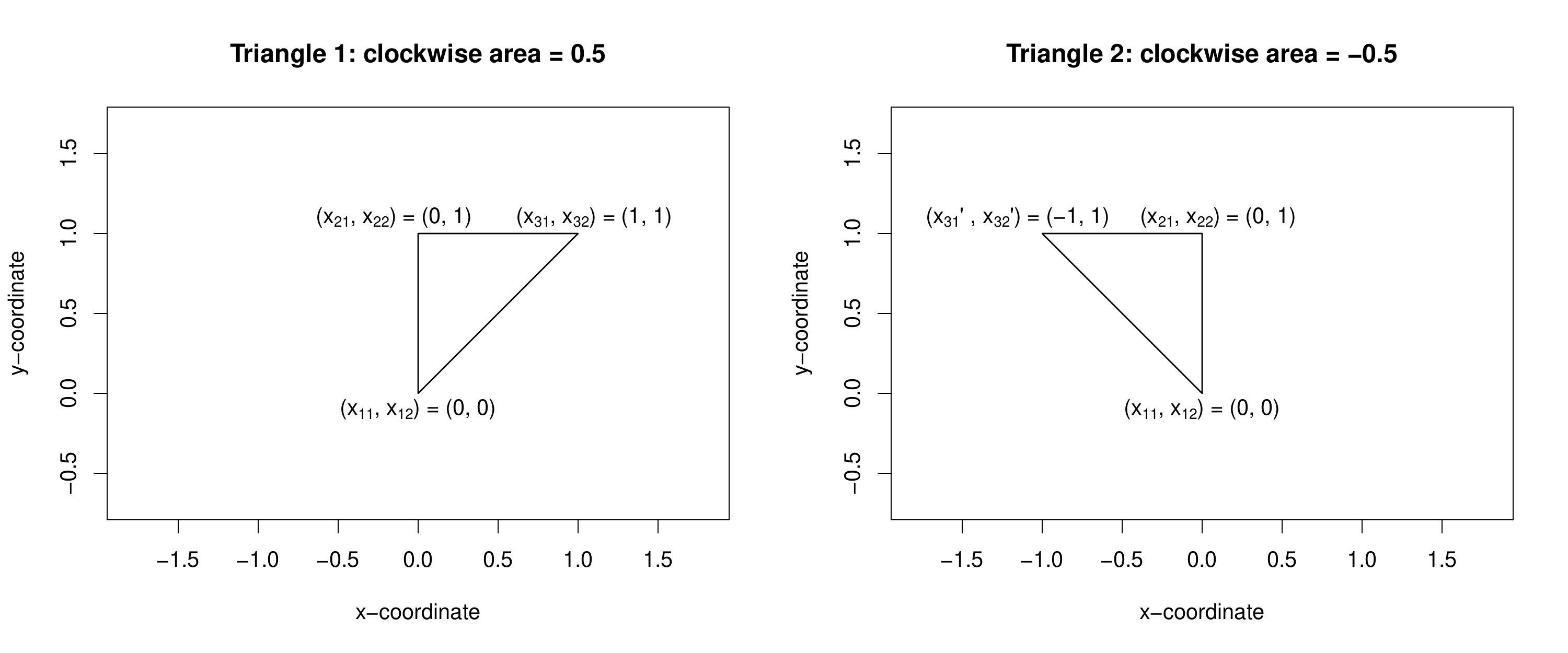}
\end{center}
\caption{\label{clockwise}Clockwise areas of triangles. Left: A triangle represented as clockwise points $(x_{11}, x_{21})$, $(x_{12}, x_{22})$, $(x_{13}, x_{23})$ with clockwise area 0.5. Right: A transformation to the left-hand triangle, in which $(x_{13}, x_{23}) \mapsto (x_{13}^*, x_{23}^*)$, giving a negative clockwise area of -0.5, based on the clockwise ordering in the left-hand triangle. Such negative areas are used to identify grids that have folded.}
\end{figure}

\begin{figure}[h!]
\begin{center}
\includegraphics[width=.98\textwidth]{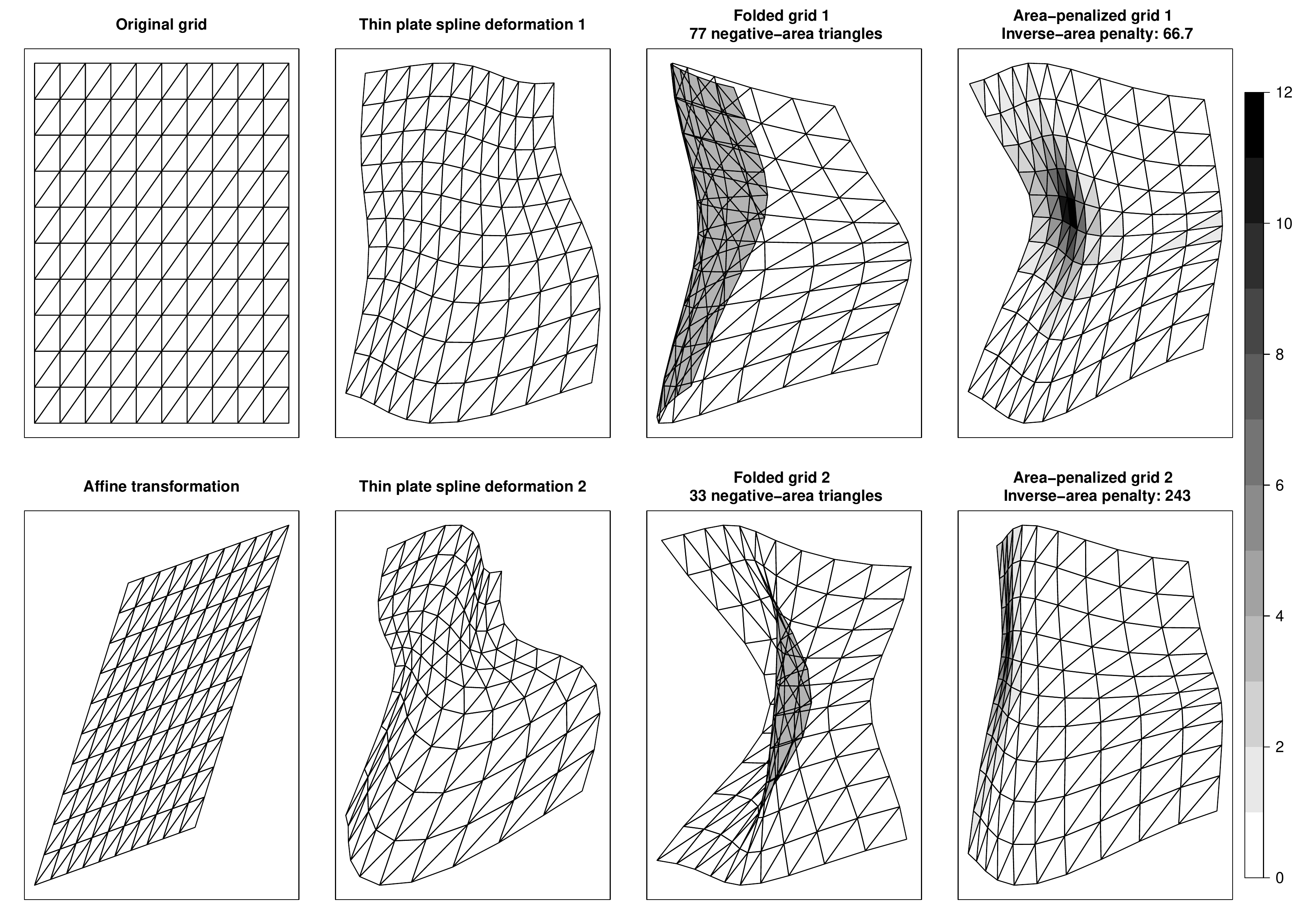}
\end{center}
\caption{\label{fold}Representations of spatial deformations using triangular tilings. Column 1, row 1: The original triangular tiling of the domain-spanning grid for calculating clockwise triangle areas. Column 1, row 2. An example of an affine transform, i.e. $x \mapsto {\bf M}_x \bx$ and $y \mapsto {\bm M}_y \bx$, for $2 \times 2$ matrices ${\bf M}_x$, ${\bf M}_y$ and $\bx = (x, y)'$. Column 2: Examples of fold-free deformations achieved by $x \mapsto g_1(x, y)$ and $y \mapsto g_2(x, y)$, where $g_1$, $g_2$ are thin plate regression splines. Column 3: As Column 2, except examples of grids that have `folded', with numbers of a triangles with negative clockwise area indicated. Column 4: As Column 2, except penalties are placed on the inverse clockwise area of triangles, for triangles with area smaller than 0.005; amounts of penalty are indicated.}
\end{figure}

A related approach to \cite{iovleff2004} is proposed here in which $\mathcal{G}$, the domain of interest, is represented as a triangular tiling. The clockwise area of each triangle is computed, which, based on Figure \ref{clockwise}, is given by $(x_{21} x_{12} + x_{31} x_{22} + x_{11} x_{32} - x_{11} x_{22} - x_{21} x_{32} - x_{31} x_{12})/2$, where $(x_{i1}, x_{i2})$, $i=1, 2, 3$, are vertices of a triangle defined in clockwise order. Subject to the triangular tiling's finite representation of $\mathcal{G}$, a change in ordering can be used to identify non-bijective $\bg$, which is equivalent to $\bg$ turning a triangle's clockwise area negative. This is illustrated in Figure \ref{clockwise} in which the left-hand triangle has clockwise area 0.5 whereas the right-hand triangle has clockwise area $-0.5$. A space represented by a triangular tiling (see Figure \ref{fold}, row 1, column 1) with a mixture of positive and negative areas must have folded; all positive areas corresponds to a fold-free space; and all negative areas corresponds to a fold-free space that has `flipped'. Flipped spaces can be eliminated without loss of generality since equivalent distances for such spaces can be achieved if the space is flipped back.

Consider the triangular tiling $\mathcal{G} = \cup_{l=1}^L \mathcal{W}_l$, where each $\mathcal{W}_l$, for $l=1, \ldots, L$, is a triangle with clockwise area $A(\mathcal{W}_l)$. For spatial deformation models only, the penalized log-likelihood can be modified to include a further penalty on folding, i.e., \begin{equation*} \label{Eq:foldpen} \ell_{p_1}(\bbeta, \blambda) = \ell_{p_0}(\bbeta, \blambda) - \delta h\big(A(\mathcal{W}_1), \ldots, A(\mathcal{W}_L)\big),\end{equation*} for some $\delta > 0$ and function $h$. The following penalties on folding are considered. 

\subsubsection*{Strict no-fold penalty}

Choosing \begin{equation} \label{nofoldeq} h_1(w_1, \ldots, w_L) = I\Big(\Big[\sum_{i=1}^LI(w_l < 0)\Big] > 0\Big)\end{equation} with $\delta$ large, e.g. $\delta = 10^6$, where $I$ is the indicator function, heavily penalizes $\ell_{p_1}(\bbeta, \blambda)$ if \emph{any} triangles have negative clockwise area. In practice, this may lead to $\ell_{p_1}(\bbeta, \blambda)$ being non-differentiable with respect to $\bbeta$; for example, the mode of $\ell_{p_1}(\bbeta, \blambda)$ could lie on the boundary of parameter space between spaces with and without folds. This would invalidate the restricted maximum likelihood (REML) approach to smoothing parameter estimation that follows in \S\ref{S:inf:reml}.

\subsubsection*{Near-fold penalties}

Differentiability of $\ell_{p_1}(\bbeta, \blambda)$ with respect to $\bbeta$ can be ensured through appropriate choice of penalty. A simple example is the inverse-area based penalty, such as $\sum_{i=1}^L w_l^{-1}$ for $w_l > 0$. In practice this performs better with a tolerance $\epsilon$ so that $\sum_{i=1}^L \max(w_l^{-1} - 1/\epsilon, 0)$ for $w_l > 0$. This penalty is illustrated in Figure \ref{fold}. 

To allow for $w_l \leq 0$, the penalty $\sum_{i=1}^L \max(\epsilon - w_l, 0) / \epsilon$ may be preferred. It is further desirable to have that $\partial h(\epsilon, \ldots, \epsilon, w_l, \epsilon, \ldots, \epsilon)/ \partial w_l \to 0$ as $w_l \nearrow \epsilon$. Hence here \begin{equation} \label{nearfold} h_2(w_1, \ldots, w_L) = \log\Big(1 + \frac{1}{\epsilon} \sum_{i=1}^L \max(\epsilon - w_l, 0)\Big)^2\end{equation} is chosen, which has the further benefit of avoiding numerically large $h_2$ for $w_l \ll 0$. 

\subsection{Smoothing parameter estimation} \label{S:inf:reml}

REML is used here to estimate smoothing parameters. This results from recognising that the penalized likelihood's penalty is proportional to the exponent of a MVN(${\bf 0}, \bS_{\bm \lambda}^-)$ distribution and then treating $\bm \beta$ as a vector of random effects integrated out by Laplace approximation. The penalized log-likelihood will be denoted $\ell_p$, which corresponds to $\ell_{p_1}$ from \S\ref{S:inf:fold} for spatial deformations if folding is penalized and to $\ell_{p_0}$ from \S\ref{S:inf:pen} otherwise. This restricted log-likelihood takes the form \[\ell({\bm \lambda}) = \ell_p(\bbeta_\blambda, \blambda) + \frac{1}{2} \log |\bS_\blambda|_+ - \frac{1}{2} \log|{\bf H}| + \frac{M_p}{2} \log(2 \pi),\] where $|{\bf H}|_+$ denotes the product of positive eigenvalues of ${\bf H}$, ${\bf H}$ is the negative Hessian of $\ell_p(\bbeta, \blambda)$ evaluated at $\hat \bbeta_\blambda$ and $M_p$ is number of zero eigenvalues in $\bS_{\bm \lambda}$. Estimating $\bm \lambda$ is an iterative procedure in which each evaluation of $\ell({\bm \lambda})$ involves estimating $\bbeta_\blambda$. Reliance on the Hessian matrix in $\ell({\bm \lambda})$ motivates the use of a twice differentiable penalty with respect to $\bbeta$ when avoiding spatial deformations folding. Where use of a full likelihood is not practical, such as if $\ell(\bbeta)$ in relation \eqref{spatial} were a composite likelihood for a max-stable process \citep{lindsay1988, padoan2010}, generalized cross-validation can be used for smoothing parameter estimation; see, e.g., \citet[Appendix A2]{rigby2005}.

\subsection{Uncertainty estimation} \label{S:meth:uncert}

Uncertainty in $D$-space can be quantified, once parameters have been estimated, through ${\bf H}$. Using Fisher information arguments relevant to penalized likelihoods, the estimated sampling distribution of $\hat{\bm \beta}_{\bm \lambda}$ is $MVN(\hat{\bm \beta}_{\bm \lambda}, {\bf H}^{-1})$, which relies on fixed $\blambda$. Smoothing parameter uncertainty can be propagated to uncertainty in $D$-spaces using the method of \citet[\S 6.11.1]{wood-book}, or the more general method of \cite{rue2009}. Examples of how such uncertainties in $D$-space can be conveyed are given in \S\ref{S:solar:uncert}.

\section{Established example: British Columbia solar radiation} \label{S:solar}

\begin{figure}[t!]
\begin{center}
\includegraphics[width=.49\textwidth, page=2]{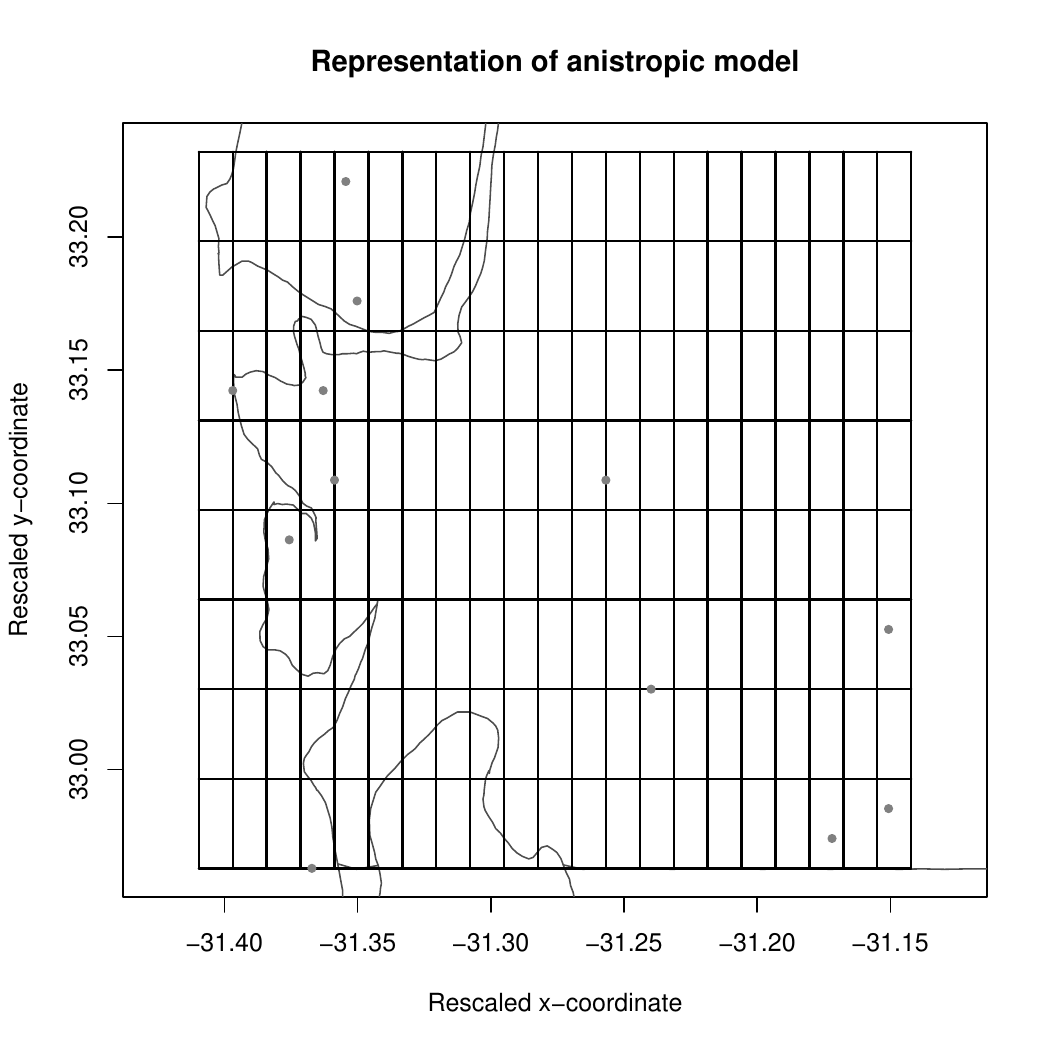}
\includegraphics[width=.49\textwidth, page=3]{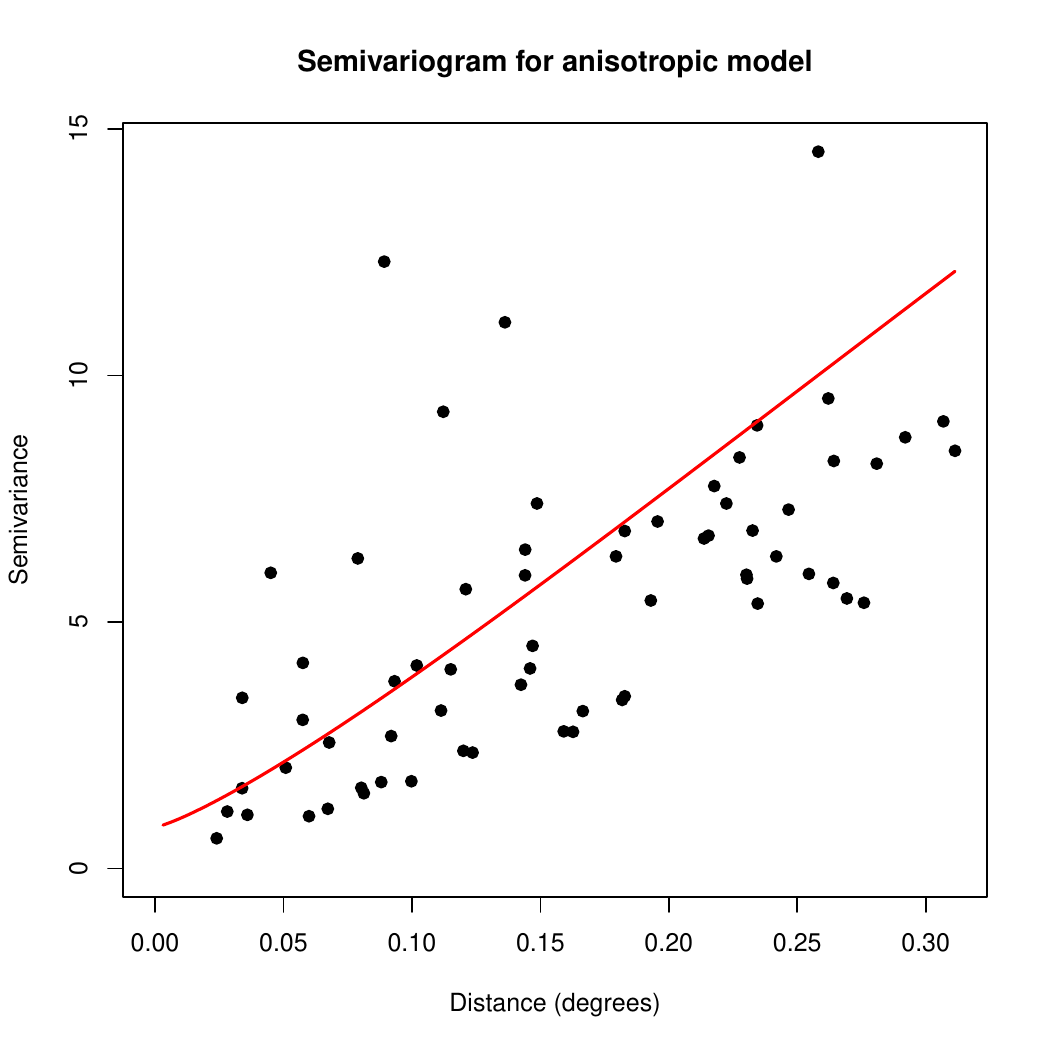}
\end{center}
\caption{\label{aniso}Conventional anisotropic model. Left: Visualisation of anisotropic grid with grey points representing station locations and outline of coast superimposed. Right: Empirical semivariogram with powered exponential model-based estimate superimposed.}
\end{figure}

The section demonstrates the methods introduced in \S\ref{meth} and \S\ref{S:inf} on solar radiation data for British Columbia. These data were used in \cite{sampson1992}'s original paper on spatial deformations, and originated from \cite{hay1984}. They are used as proof-of-concept data, due to their popularisation in subsequent related works, such as \cite{schmidt2003} and \cite{bornn2012}. Similarly to \cite{schmidt2003}, the spring-summer dataset is studied here, which comprises $T=732$ measurements (22 March 1980 -- 20 September 1983) on solar radiation at 12 monitoring stations. The resulting semivariogram for the data is shown in Figure \ref{aniso}. The semivariogram shows clear deviation from a monotonic relationship between the estimated semivariances and distance. 

Both spatial deformation and dimension expansion models are considered for these data. The data are de-trended, as in \cite{sampson1992}, scaled to have zero mean at each station, and then modelled as a Gaussian process using their empirical $12 \times 12$ covariance matrix, ${\bm V}$. Inference is therefore based on the likelihood given in \eqref{gplik}, where ${\bm \Sigma}$ has elements $\Sigma_{ij} = \gamma(||\bg(\bx_i) = \bg(\bx_j)||; \cdot)$. Models will be compared against a conventional anisotropic model: i.e.,  $\bx^* = (x_1 / \phi_1, x_2 / \phi_2)$, where $\phi_1, \phi_2 > 0$ are scale parameters, in the notation of \S\ref{S:meth:nonstat}. This work considers only the powered exponential covariance function, given by \begin{equation*} \gamma(h; \sigma^2, \tau^2, \alpha) = \left\{\begin{array}{ll} \sigma^2 + \tau^2 & \text{if  } h = 0,\\ \sigma^2 \exp(-h ^\alpha) & \text{otherwise,}\end{array}\right. \end{equation*} for $0 < \alpha \leq 2$, due it its greater flexibility than the exponential form and greater analytical tractability than the Mat\'ern form, which is often used for environmental applications. A grid (which is the same throughout this section) is used to represent the deformation offered by the anisotropic model. This is shown in Figure \ref{aniso} alongside its model-based semivariogram superimposed on the empirical semivariogram.

\subsection{Spatial deformation}

Given basis representation \eqref{basis}, $g_1$ and $g_2$ are each chosen as rank-12 thin-plate regression splines, as introduced in \S\ref{S:meth:basis}.

\subsubsection{Folding unconstrained}

\begin{figure}[t!]
\begin{center}
\includegraphics[width=.49\textwidth, page=1]{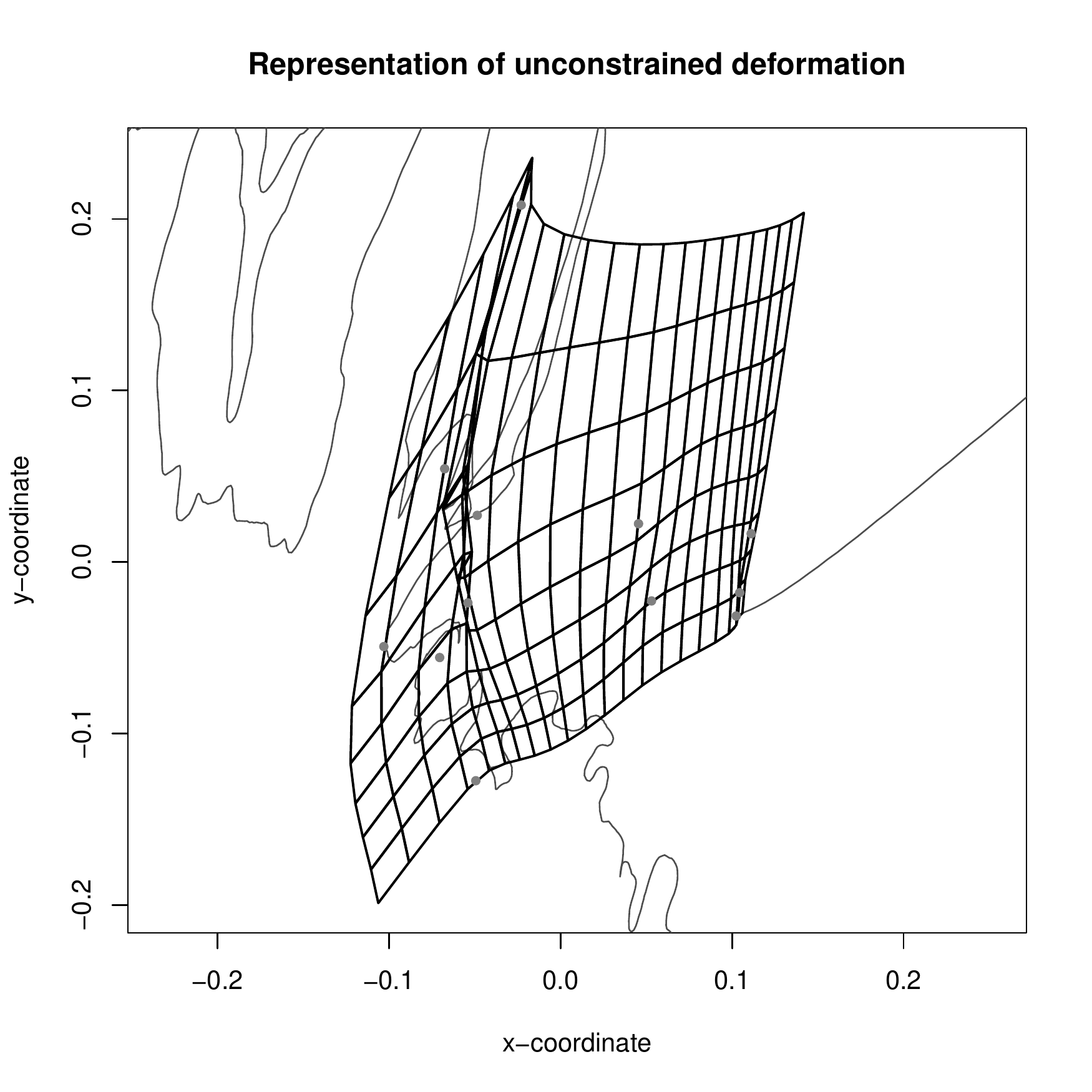}
\includegraphics[width=.49\textwidth, page=4]{aniso.pdf}
\end{center}
\caption{\label{sd1}Unconstrained spatial deformation model. Plots as described in Figure \ref{aniso}.}
\end{figure}

The first spatial deformation model fitted involves no penalty on whether $D$-space folds. A representation of the resulting deformation is shown in Figure \ref{sd1}. This shows how a regular $0.05 \times 0.05$ degree grid, previously used in Figure \ref{aniso}, is changed in $D$-space. Changes to station locations and a coastline outline are also shown. Figure \ref{sd1} also shows a semivariogram, with distances now calculated over $D$-space.

The deformed space is clearly different from that offered by the anisotropic model (Figure \ref{aniso}), which is evident from changes to the 0.05 degree grid and the semivariograms: the former is clearly not achievable by a simple scaling $G$-space in either direction and the latter shows empirical semivariances much closer to the assumed powered exponential form. Allowing for slightly different data and presentation methods, these results are consistent with the analyses of \cite{sampson1992}, \cite{schmidt2003} and \cite{bornn2012}. 

\subsubsection{Folding penalized} \label{solar-pen}

\begin{figure}[h!]
\begin{center}
\includegraphics[width=.49\textwidth, page=1]{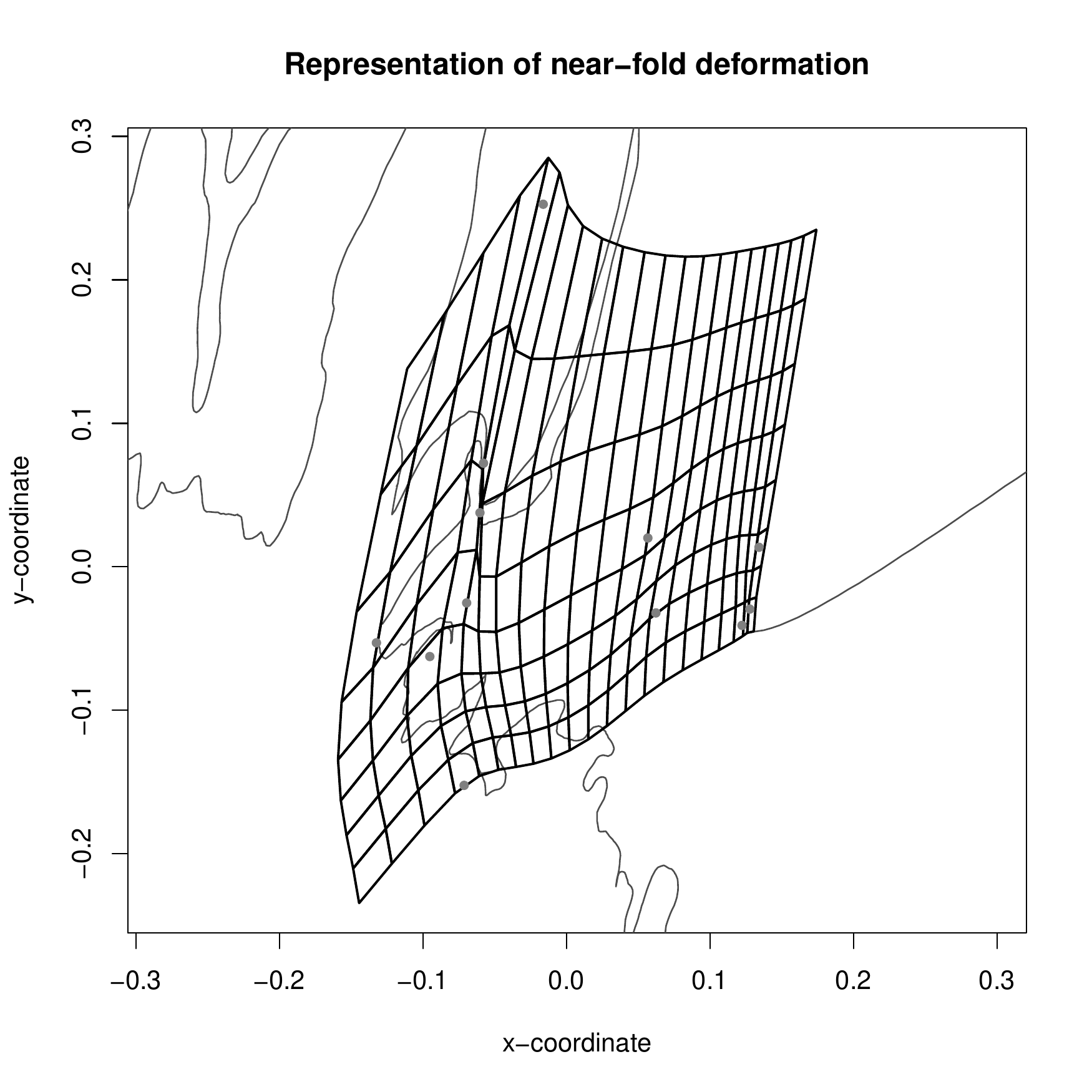}
\includegraphics[width=.49\textwidth, page=4]{penfold.pdf}
\end{center}
\caption{\label{sd3}Near-fold penalized spatial deformation model. Plots as in Figure \ref{aniso}.}
\end{figure}

Figure \ref{sd1} shows $D$-space to have folded. Following \S\ref{fold}, $D$-spaces in which this happens can be avoided. Here that is achieved by adopting the penalty of \eqref{nearfold}, taking $\epsilon = 0.1 A_\text{aniso}$ and $\delta = 10^{6}$, where $A_\text{aniso}$ represents the area of cells in the conventional anisotropic model. (Results of the strict no-fold penalty, defined in \eqref{nofoldeq}, are suppressed as they are qualitatively the same of those of this section, and because parameter estimates lie on the non-differentiable boundary separating a bijective and non-bijective $D$-space, which invalidates \S\ref{S:inf:reml}.) The representation of $D$-space in Figure \ref{sd3} shows the fold of Figure \ref{sd1} to have gone, while the remainder of $D$-space remains essentially unchanged. The effect on the semivariogram caused by applying the penalty of \eqref{nearfold} appears minimal.

The effects of the different approaches to folding, ranging from allowing to discouraging, are as expected. Performing objective inference on basis coefficients and smoothing parameters when folding is unconstrained is relatively straightforward. Penalising $D$-spaces that are near folding, in particular the parameters that control the penalty if according to \eqref{nearfold}, is rather more subjective. Here $\epsilon$ has been specified in terms of $A_\text{aniso}$, which partly negates effects of differing domain sizes. Further discussion of parameter choice when penalizing near-folding is given in \S\ref{discuss}.

\subsection{Dimension expansion} \label{S:solar:dimexp}

Now one- and two-dimensional dimension expansion models are considered, which refers to the number of \emph{added} dimensions.

\subsubsection{One-dimensional expansion} \label{solar-1d}

For a one-dimensional dimension expansion, $\bx = (x_1, x_2) \mapsto \bx^* = (x_1/\phi, x_2/\phi, z_1)$ for $\bx \in \mathcal{S}$, where $z_1 = g_1(\bx)$. Here $g_1$ is chosen as a rank-12 thin plate regression spline and $\phi$ is formulated as a basis coefficient, as described in \S\ref{S:inf:pen}. The estimated additional dimension is shown in Figure \ref{dim1} for the study domain alongside the semivariogram with distance based on $\bx^*$ in three-dimensional $D$-space.

\begin{figure}[h!]
\begin{center}
\includegraphics[width=.53\textwidth, page=1]{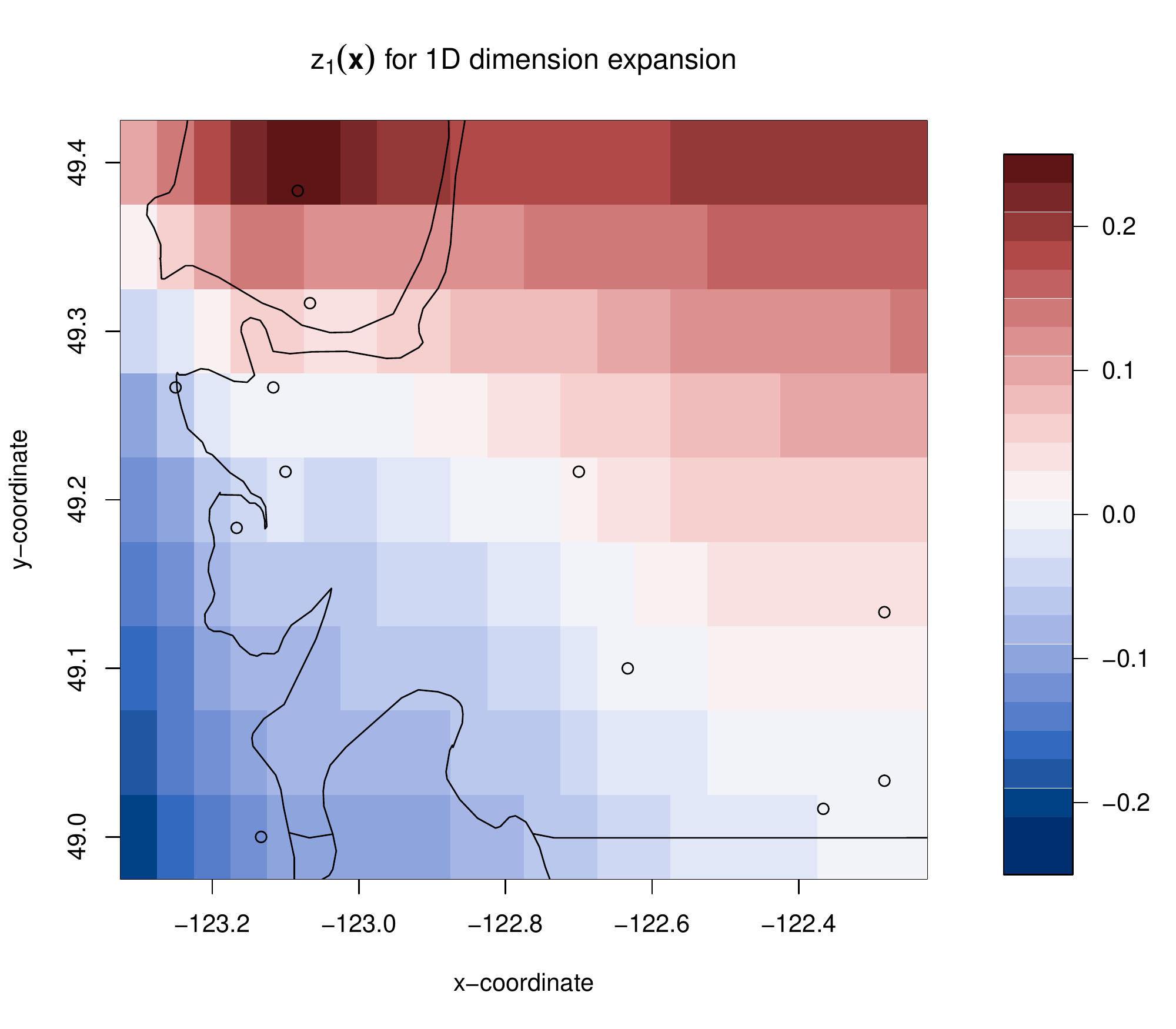}
\includegraphics[width=.46\textwidth, page=2]{dim1-2.pdf}
\end{center}
\caption{\label{dim1}One-dimensional dimension expansion model. Left: Representation of $z_1  = g_1(\bx)$, the added dimension. Right: Empirical semivariogram with powered exponential model-based estimate superimposed and distance based on $\bx^*$ in three-dimensional $D$-space.}
\end{figure}

Comparison between the spatial deformation and dimension expansion estimates is perhaps not immediate. It is first worth restricting attention to convex hull determined by the station locations. Then consider the northernmost point: the Grouse Mountain station. This was singled out in the previous analyses of \cite{sampson1992} and \cite{schmidt2003}, primarily for its elevation, which, at 1128m, is notably higher than 125m, the height of the next highest station. Its $D$-space representation is consistent between the spatial deformation and dimension expansion models, once an overall scaling is taken into account: in $D$-space in the former it is further away from the remaining points than in $G$-space, and in the latter its corresponding new dimension is the point most different from zero, which separates it most from the other points in comparison to $G$-space. Conversely, for the spatial deformation model, those points located in $D$-space where grid cells have smallest area correspond to the points of the added dimension that are close to zero in the dimension expansion model. The model's semivariogram based on $D$-space is perhaps most like the spatial deformation model in which folding was allowed. However, as the dimension expansion model requires only one as opposed to two rank-12 thin plate regression splines, it therefore has 12 fewer parameters (12 basis coefficients fewer, one smoothing parameter fewer, but an additional unknown $\phi$). 

\subsubsection{Two-dimensional expansion} \label{solar-2d}

A two-dimensional dimension expansion model is now fitted with each dimension represented by a rank-12 thin plate regression spline. This model requires 13 more parameters than its one-dimensional counterpart (12 basis coefficients and one smoothing parameter). 

\begin{figure}[h!]
\begin{center}
\includegraphics[width=.99\textwidth, page=1]{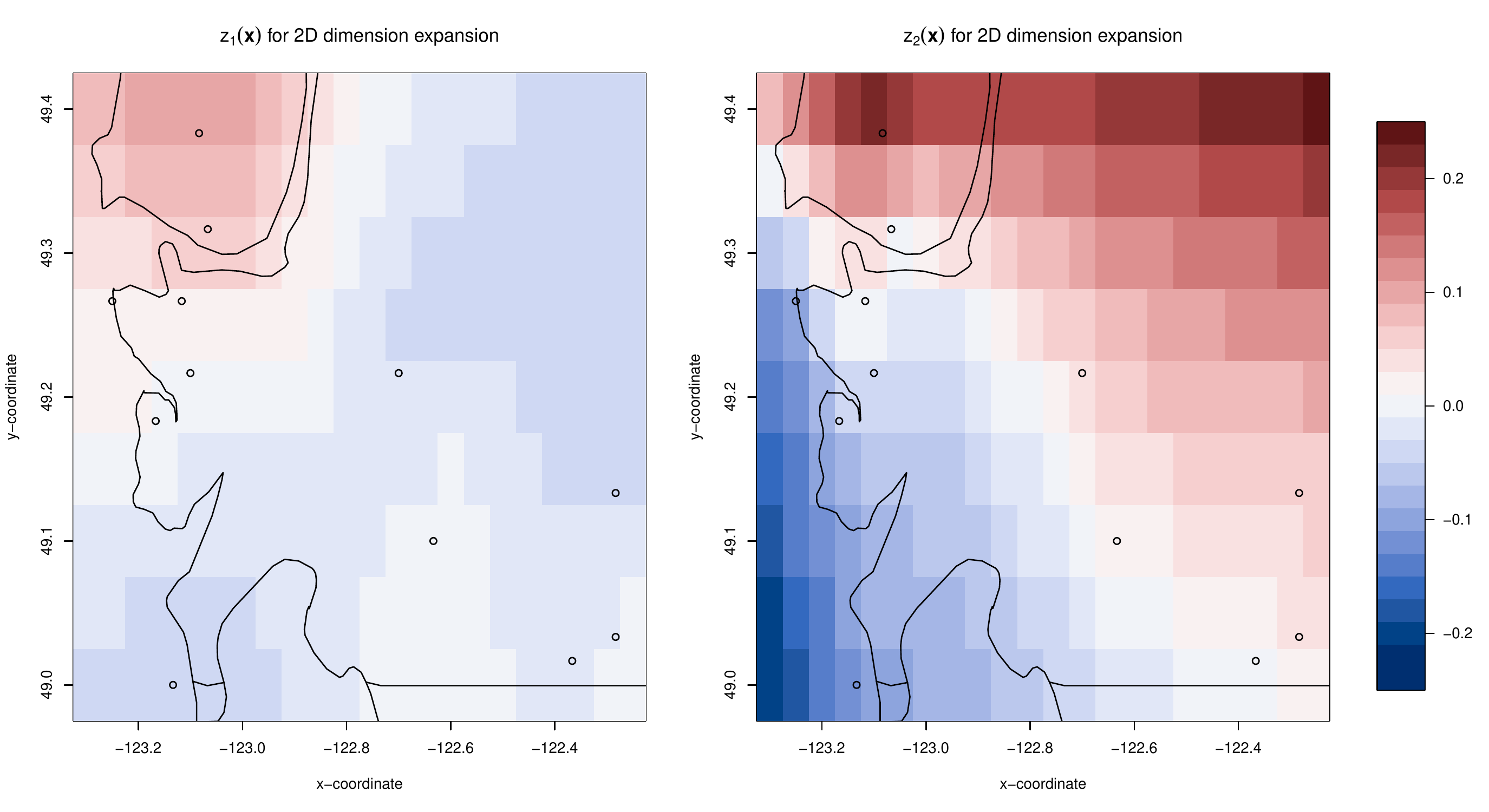}\\
\includegraphics[width=.53\textwidth, page=1]{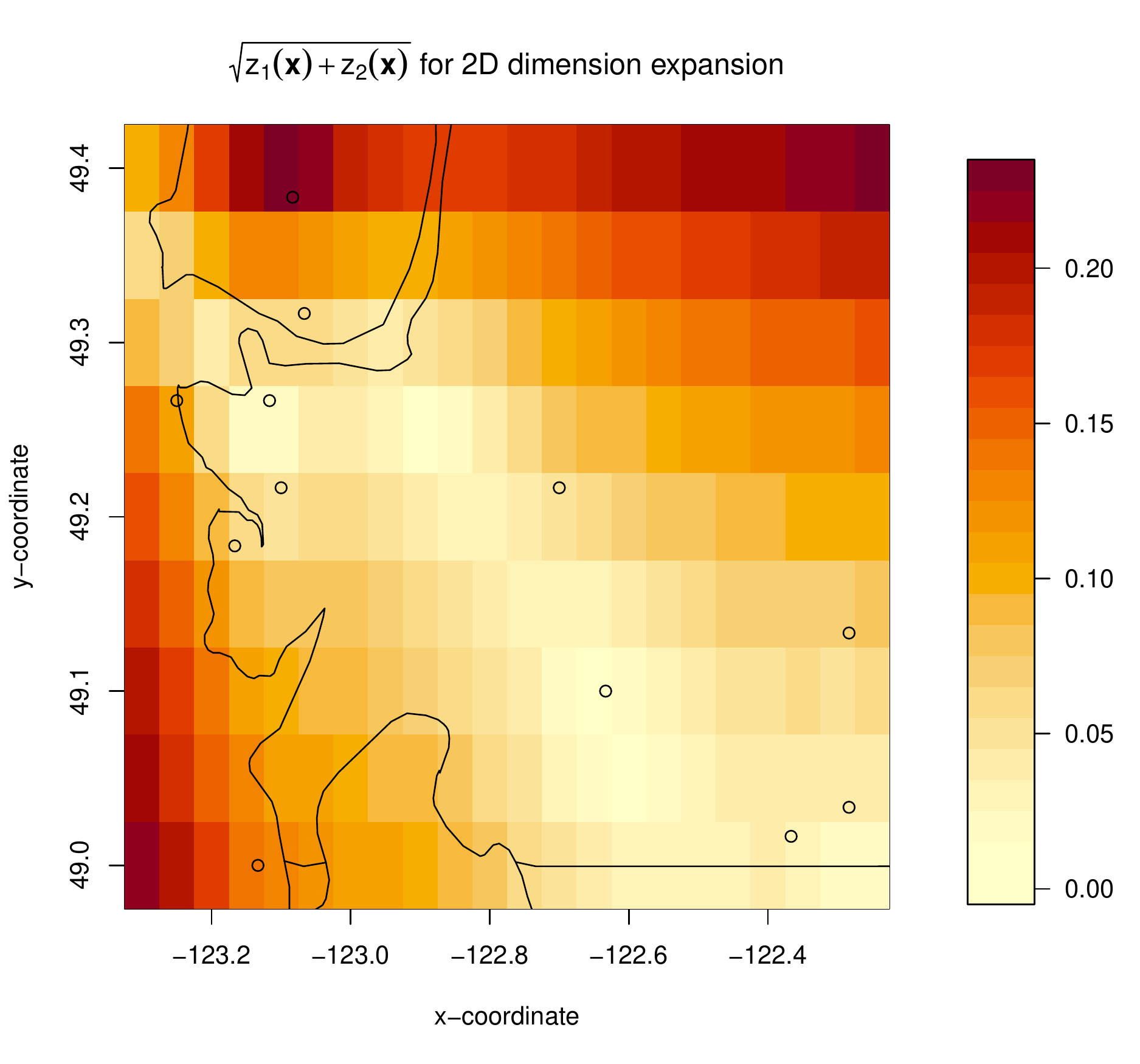}
\includegraphics[width=.46\textwidth, page=2]{dim2-3.pdf}
\end{center}
\caption{\label{dim1}Two-dimensional dimension expansion model. Row 1: separate representations of added dimensions, $z_1  = g_1(\bx)$ and $z_2  = g_2(\bx)$. Row 1, column 1: combined representation of added dimensions, $z_1^2 + z_2^2$. Row 1, column 2: Empirical semivariogram with powered exponential model-based estimate superimposed and distance based on $\bx^*$ in four-dimensional $D$-space.}
\end{figure}

The results of adding a dimension in the dimension expansion approach seem to follow naturally from the one-dimensional model. The second dimension in the two-dimensional model closely resembles that of the one-dimensional model, allowing for negated $z_1$ values, to which the covariance structure is invariant. The first dimension appears near-zero across the domain except for around the Grouse Mountain station, which is further separated from the other stations by the additional dimension. Points on the empirical semivariogram appear to lie closer to the powered exponential model-based estimate for the two-dimensional dimension expansion model compared to its one-dimensional counterpart, but the difference is relatively small. Formal testing should be considered for choosing an optimal number of dimensions: see \S\ref{discuss}.

\subsection{Uncertainty estimates} \label{S:solar:uncert}

\begin{figure}[h!]
\begin{center}
\includegraphics[width=.99\textwidth, page=1]{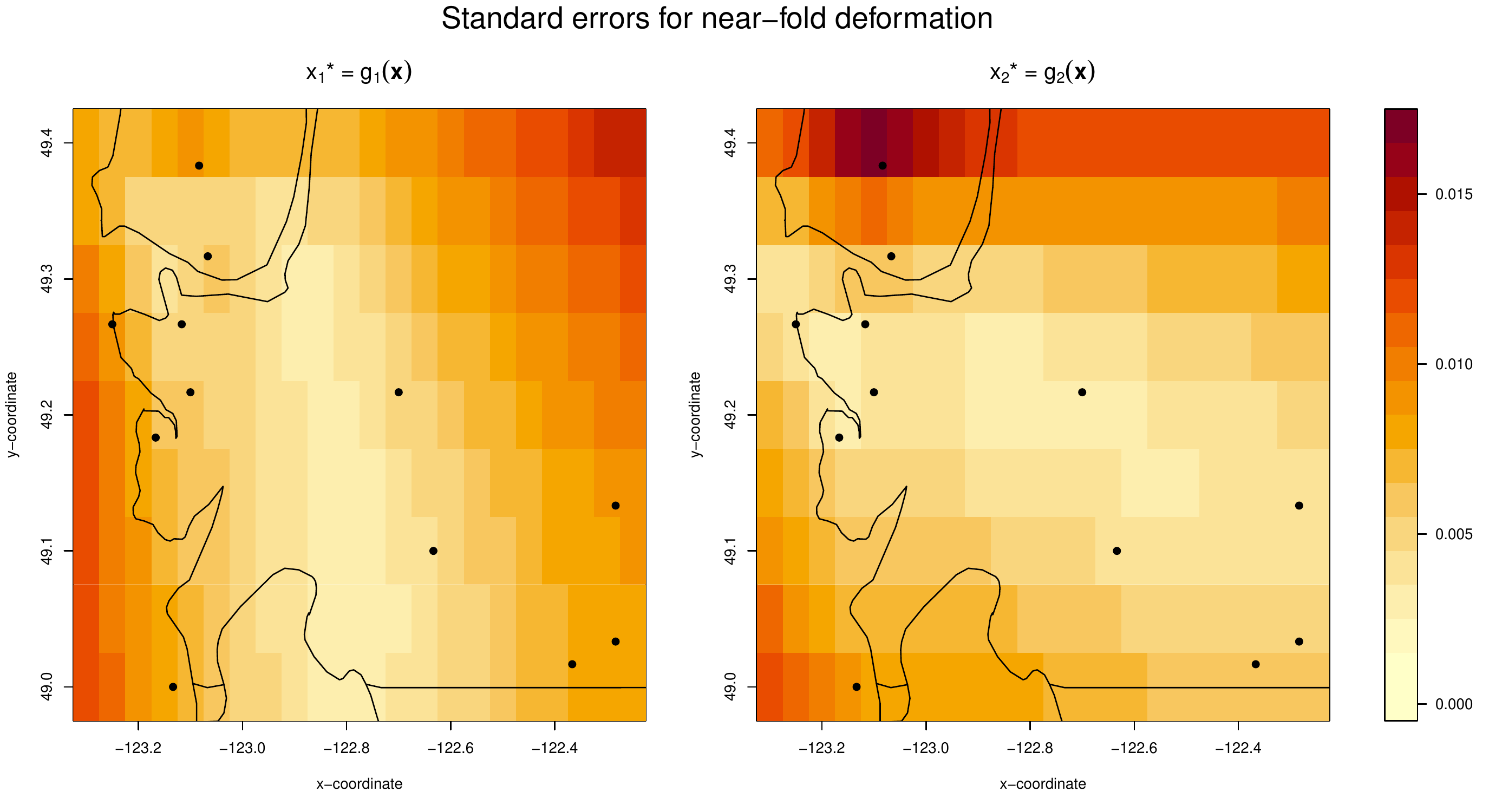}
\end{center}
\caption{\label{unc1}Standard errors for coordinate in $D$-space, given coordinates in $G$-space, for the spatial deformation model of \S\ref{solar-pen} in which folding is prohibited.}
\end{figure}

Uncertainty estimates for the spatial deformation model in which near-folding is penalized are shown in Figure \ref{unc1}. These are represented by standard errors of $D$-space coordinates given $G$-space coordinates.

In general standard errors are seen to be smaller nearest the stations and grow as stations become more distance. The exception for this, for both $D$-space coordinates, is the Grouse Mountain station, which has largest standard errors. This is likely to be a consequence of its location in $D$-space being most transformed in comparison to the other stations.

\begin{figure}[h!]
\begin{center}
\includegraphics[width=.99\textwidth, page=1]{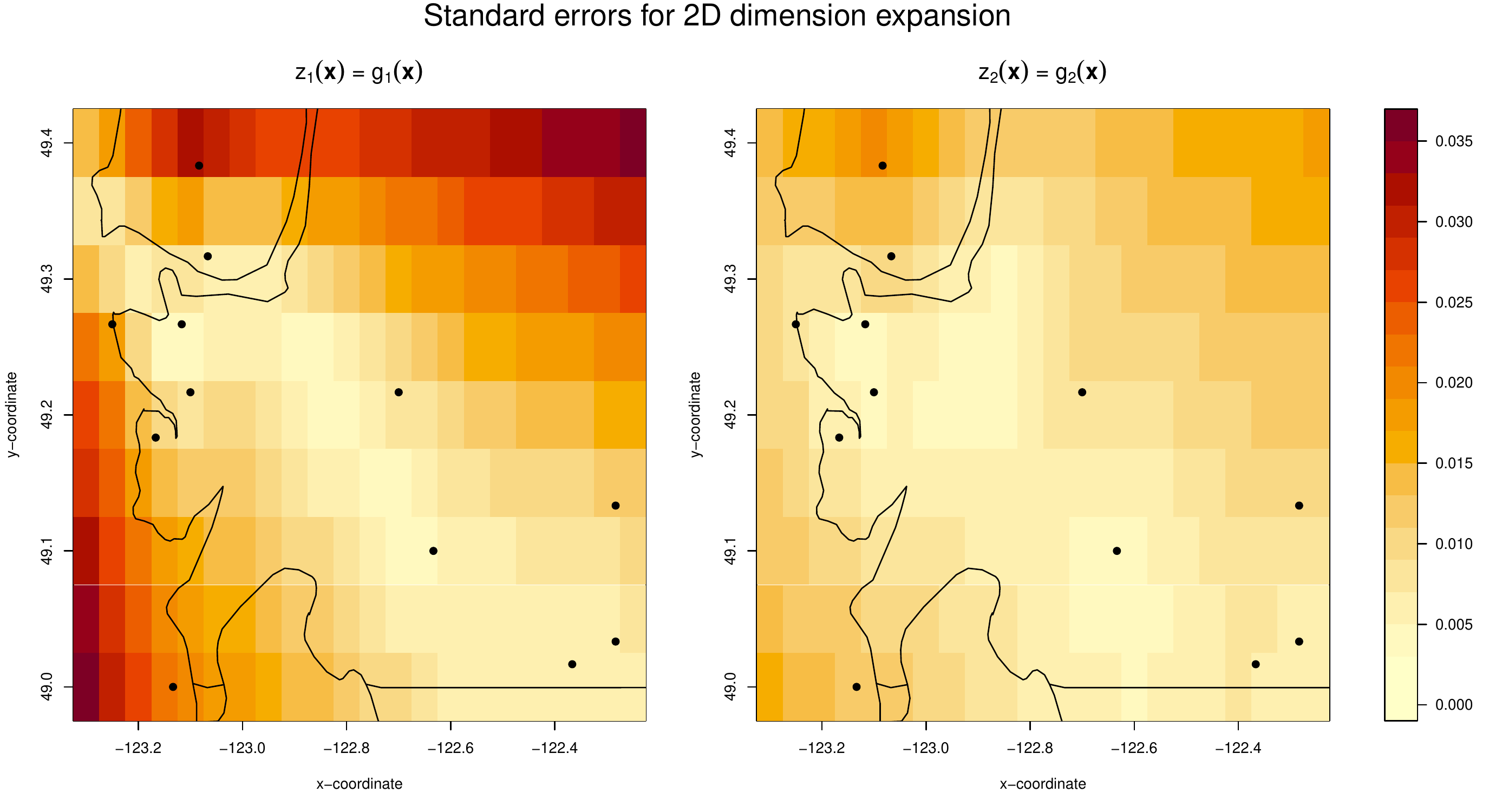}
\end{center}
\caption{\label{unc2}Standard errors for dimensions added to $D$-space, given coordinates in $G$-space, for the dimension expansion model of \S\ref{solar-2d} in which two dimensions are added.}
\end{figure}

Figure \ref{unc2} shows standard errors for each added dimension in the two-dimensional dimension expansion model of \S\ref{solar-2d}. Variation of standard errors with coordinates in $G$-space is similar to that of the spatial deformation model. Closer inspection suggests that their increase as stations become more distant has a slightly greater effect compared to their inflation for Grouse Mountain in comparison to standard errors in the spatial deformation model. Note also that direct comparison of standard errors between the spatial deformation and dimension expansion models is not immediately possible due to the effect of $\phi$ in the latter.

\section{Risk modelling: Extreme Colorado rainfall} \label{S:rain}

This section presents an analysis relevant to risk estimation by developing a model that can simulate extreme daily rainfall accumulations.

\subsection{Data}

\begin{figure}[h!]
\begin{center}
\includegraphics[width=.99\textwidth]{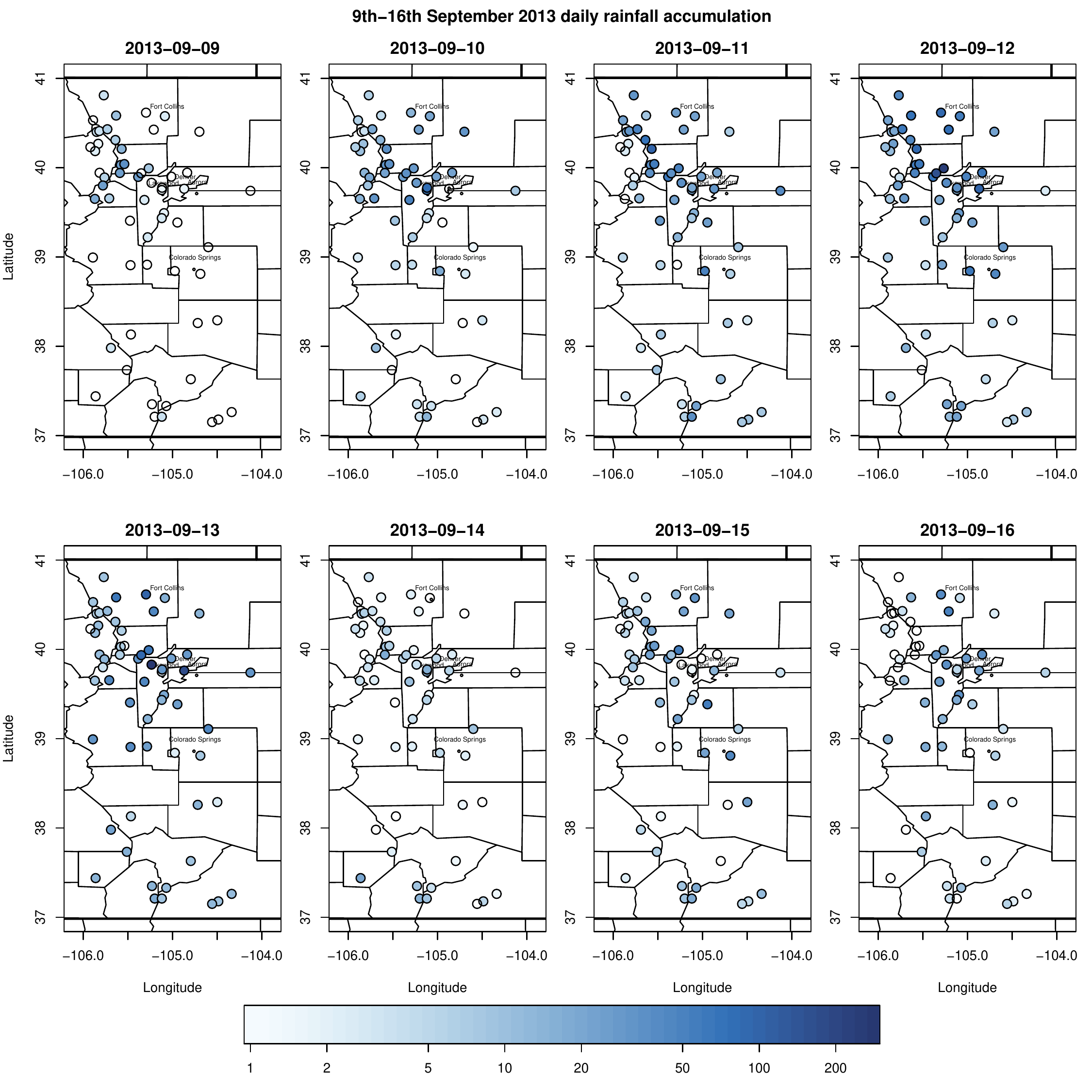}
\end{center}
\caption{\label{flood}Daily rainfall accumulations (mm) for part of Colorado, US, between 9th and 16th September 2013, i.e., spanning the 2013 Colorado floods.}
\end{figure}

Extreme rainfall data over part of Colorado from 1st April to 31st October are studied. This region and time range were originally chosen by \cite{cooley2007}. Figure \ref{flood} shows daily rainfall accumulations from 9th to 16th September 2013 over the study region, which covers the 2013 Colorado Floods. The heavy rainfall amounts on 11th, 12th and 13th September 2013 are particularly prominent.

\subsection{Marginal model}

Extreme daily rainfall is defined locally as exceeding a threshold $u({\bm x})$, which is estimated as the $100(1 - \zeta)^\text{th}$ percentile of daily rainfall. Here $\zeta = 0.03$ is chosen. As estimates of this percentile and the distribution of its excesses are required for every location in the study region, a spatially continuous approach is used. This is achieved, following \cite{y2019}, using generalized additive model forms for distribution parameters. The threshold is then estimated by quantile regression, through the asymmetric Laplace distribution (ALD), and its excesses modelled as realisations from the generalized Pareto distribution (GPD). Let $Y_t({\bm x})$ denote the daily rainfall accumulation at location ${\bm x}$ in region $\mathcal{G}$ at time $t=1, \ldots, T$. Specifications of the GPD and quantile regression models are given respectively by \[Y_t({\bm x})  - u({\bm x}) \mid Y_t({\bm x}) > u({\bm x}) \sim GPD(\psi({\bm x}), \xi({\bm x}))\] where \begin{align*}\log \psi({\bm x}) &= \beta_\psi + f_{\psi, \text{tp}}({\bm x}) + f_{\psi, \text{cr}}\big(elev({\bm x})\big)\\ \label{gpd2} \xi({\bm x}) &= \beta_\xi + f_{\xi, \text{tp}}({\bm x}) + f_{\xi, \text{cr}}\big(elev({\bm x})\big) \end{align*}  where \[Y_t({\bm x}) \sim ALD(u({\bm x}), \sigma({\bm x}))\] with \begin{align*}u({\bm x}) &= \beta_u + f_{u, \text{tp}}({\bm x}) + f_{u, \text{cr}}\big(elev({\bm x})\big)\\ \label{ald2} \log \sigma({\bm x}) &= \beta_\sigma + f_{\sigma, \text{tp}}({\bm x}) + f_{\sigma, \text{cr}}\big(elev({\bm x})\big). \end{align*} In the above equations $f_{*, \text{tp}}$ and $f_{*, \text{cr}}$ denote thin plate and cubic regression splines, respectively, and $elev({\bm x})$ denotes the elevation of location ${\bm x}$. 

\begin{figure}[h!]
\begin{center}
\includegraphics[width=.32\textwidth, page=1]{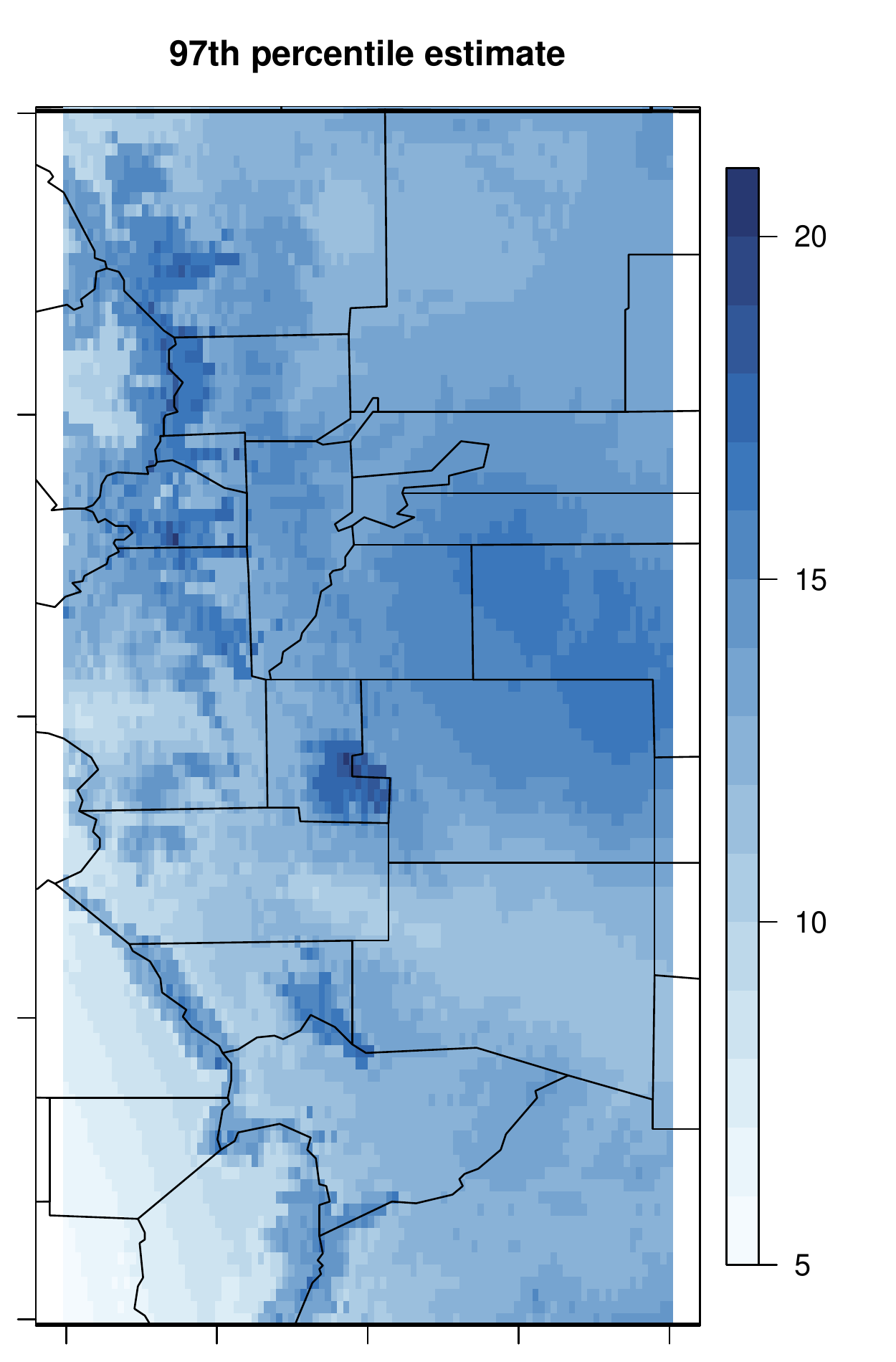}
\includegraphics[width=.32\textwidth, page=2]{pars.pdf}
\includegraphics[width=.32\textwidth, page=3]{pars.pdf}
\end{center}
\caption{\label{F:rain:pars}Colorado daily rainfall: The exceedance threshold, i.e., the 97th percentile estimate, and GPD scale and shape parameter estimates based on exceedances of threshold.}
\end{figure}

Estimates of the threshold (i.e. the 97th percentile), $u({\bm x})$, and the GPD scale, $\psi({\bm x})$, and shape, $\xi({\bm x})$, parameters are shown in Figure \ref{F:rain:pars}. The threshold estimates clearly shows an increase with elevation, whereas the scale parameter decreases with elevation. The latter relationship is qualitatively similar to that in \cite{cooley2007}, although direct comparison is not possible due to differences in threshold and GPD scale parameter specifications. 

\subsection{Spatial model}

The threshold and GPD models allow spatially continuous probability integral transformations of rainfall measurements to an arbitrary scale. Here measurements are transformed to unit Gaussian scale so that the joint distribution of transformed rainfall over space can be modelled as a Gausian process. Transforming to unit Frech\'et scale and modelling using a max-stable process was considered first; see, e.g., \cite{dav-ghol}, \cite{dav-sum}. However, a model that imposes asymptotic independence \citep{sibuya1960, coles1999} was seen to be more appropriate; hence a Gaussian process model is used here.

Due to a large proportion of zeros in the Colorado rainfall data, no sensible probability integral transformation can yield approximately Gaussian data. Furthermore, as the aim here is to simulate extreme rainfall, it is most important to capture dependence between extreme values, as opposed to lesser values. Consequently, a \emph{tail} Gaussian process is fitted via the tail bivariate Gaussian model of \cite{bor-ct}. Put simply, this involves treating non-exceedances of the threshold as censored. The tail Gaussian process could be fitted by through its full likelihood. However this involves evaluating the multivariate Gaussian distribution's cumulative distribution function at all locations where data are censored, and for each time point, which is computationally prohibitive even for moderate numbers of time points and/or locations. A slightly less accurate---but typically much quicker---approach is to consider all pairwise combinations of locations, estimate their covariances through the tail bivariate Gaussian model of \cite{bor-ct}, and then use these to populate the covariance matrix for all sites. This enables inference through likelihood \eqref{gplik}. There are some scenarios of model and/or data scenarios for which this simplification is unsuitable, some of which are discussed in \S\ref{discuss}.

Let $F_{GPD}(\,; \psi({\bm x}), \xi({\bm x}))$ denote the GPD cumulative distribution function (cdf) with scale and shape parameters at location ${\bm x}$, $\psi({\bm x})$ and $\xi({\bm x})$, respectively. Margins are converted to Gaussian, for exceedances of the threshold, through the probability integral transformation given by \begin{equation} \label{E:pit} \text{   } \hspace{1cm} Z_t({\bm x}) = \left\{\begin{array}{ll} \Phi_1^{-1}\Big(1 - \zeta \big[1 - F_{GPD}\big(Y_t({\bm x}) - u({\bm x}); \psi({\bm x}), \xi({\bm x})\big)\big]; 0, 1\Big) & \text{if } Y_t({\bm x}) > u({\bm x}),\\ \Phi_1^{-1}(1 - \zeta; 0, 1) & \text{if } Y_t({\bm x}) \leq u({\bm x}),\end{array}\right.\end{equation} where $\Phi_1(\, ; \mu, \tau^2)$ denotes the mean $\mu$ variance $\tau^2$ Gaussian cdf with inverse $\Phi_1^{-1}(\, ; \mu, \tau^2)$ and probability density function (pdf) $\phi_1(\, ; \mu, \tau^2)$. 

Now consider populating the sample covariance matrix for a finite set of locations ${\bm x}_1, \ldots, {\bm x}_D$. Given the conversion of margins to unit Gaussian here, a sample correlation matrix may be populated. Let ${\bm z}({\bm x}) = (z_1({\bm x}), \ldots, z_T({\bm x}))$ denote the realizations for location ${\bm x}$. For each pair of locations, ${\bm x}_i$ and ${\bm x}_j$, say, for $i \neq j$, let \[\begin{pmatrix} Z_t({\bm x}_i)\\ Z_t({\bm x}_j) \end{pmatrix} \mid \rho_{ij} \sim BVN\left(\begin{pmatrix} 0\\ 0 \end{pmatrix}, \begin{pmatrix} 1 & \rho_{ij}\\ \rho_{ij} & 1 \end{pmatrix}\right)\] and denote the bivariate standard bivariate Gaussian pdf and cdf with correlation $\rho$ by  $\phi_2(\, , \, ; \rho)$ and $\Phi_2(\, , \, ; \rho)$, respectively. For $i < j$, a maximum likelihood estimate of $\rho_{ij}$, $\hat \rho_{ij}$, is found by maximising likelihood \[L\big(\rho_{ij}; {\bm z}({\bm x}_i), {\bm z}({\bm x}_j)\big) = \prod_{t=1}^T f\big(z_t({\bm x}_i), z_t({\bm x}_j); \rho_{ij}\big)\] with respect to $\rho_{ij}$, where $f\big(z_t({\bm x}_i), z_t({\bm x}_j); \rho_{ij}\big)$ is given by \begin{equation*}\left\{\begin{array}{lll} \Phi_2\big(z_t({\bm x}_i), z_t({\bm x}_j); \rho) & \text{if} & y_t({\bm x}_i) \leq u({\bm x}_i), y_t({\bm x}_j) \leq u({\bm x}_j),\\ \Phi_1(z_t({\bm x}_i); \rho z_t({\bm x}_j); 1 - \rho^2) \phi_1(z_t({\bm x}_j); 0, 1) & \text{if} & y_t({\bm x}_i) \leq u({\bm x}_i), y_t({\bm x}_j) > u({\bm x}_j),\\ \Phi_1(z_t({\bm x}_j); \rho z_t({\bm x}_i), 1 - \rho^2) \phi_1(z_t({\bm x}_i); 0, 1) & \text{if} & y_t({\bm x}_i) > u({\bm x}_i), y_t({\bm x}_j) \leq u({\bm x}_j),\\ \phi_2\big(z_t({\bm x}_i), z_t({\bm x}_j); \rho) & \text{if} & y_t({\bm x}_i) > u({\bm x}_i), y_t({\bm x}_j) > u({\bm x}_j).\end{array}\right. \end{equation*}

\begin{figure}[h!]
\begin{center}
\includegraphics[width=.49\textwidth]{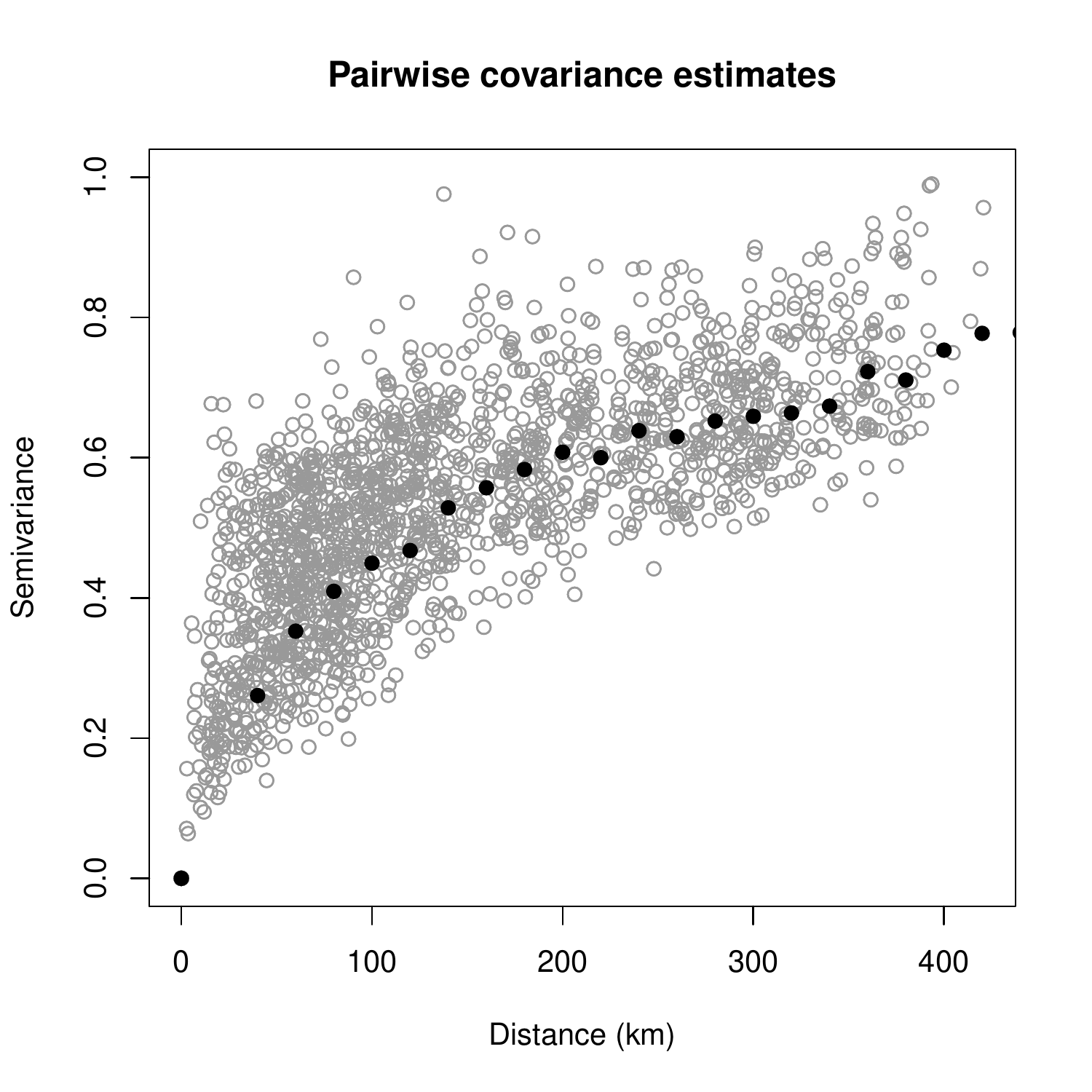} \includegraphics[width=.49\textwidth, page=1]{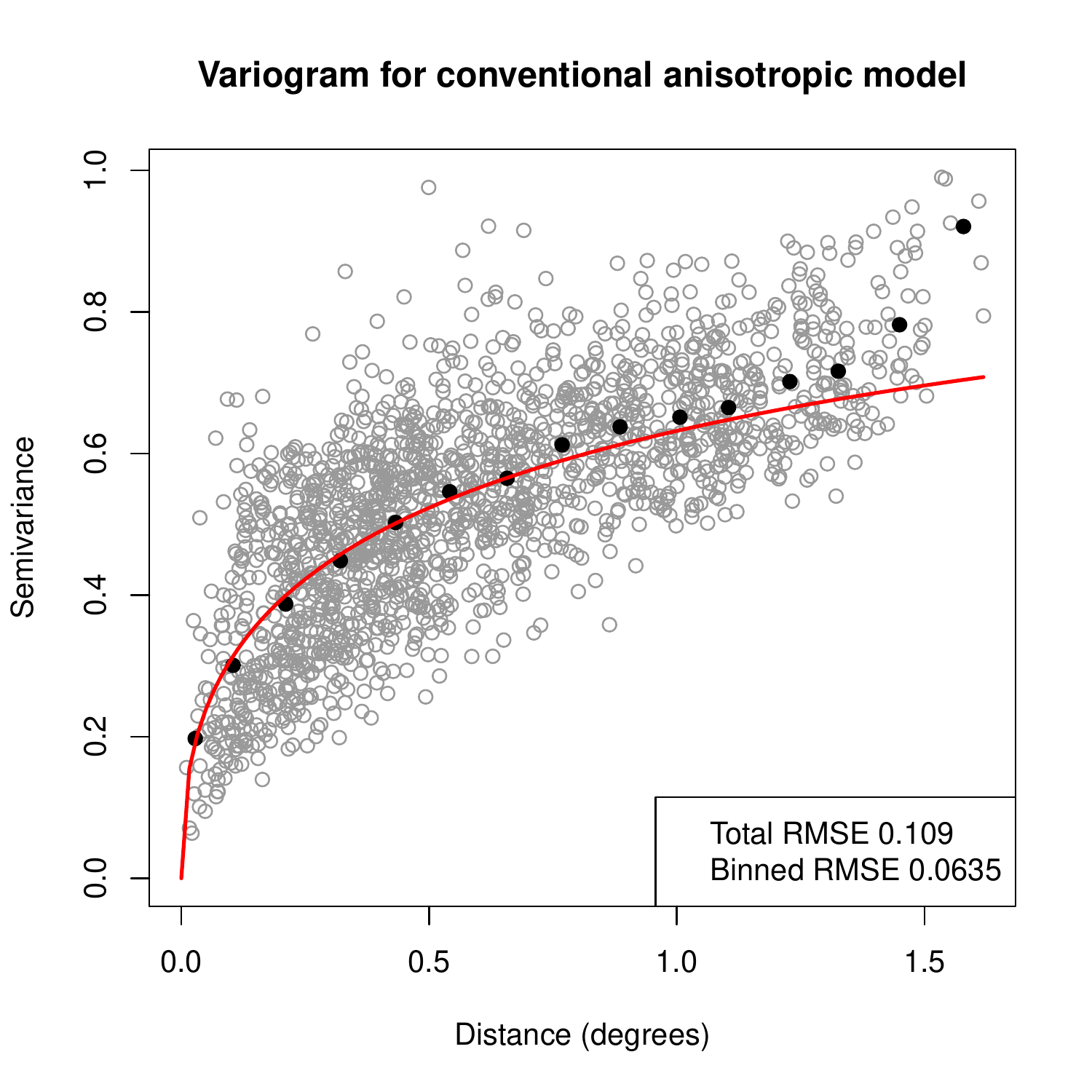}
\end{center}
\caption{\label{F:rain:iso}Binned semivariance based on conventional anisotropic and deformed geostatistical models.}
\end{figure}

Semivariance estimates, derived from $\hat \rho_{ij}$, are shown against great circle distance in Figure \ref{F:rain:iso}. These show a general increase in semivariance with distance. Figure \ref{F:rain:iso} also shows semivariance estimates against distance in degrees based on fitting a conventional anisotropic Gaussian process, i.e. if ${\bm x} = (x_1, x_2) \mapsto {\bm x}^* = (x_1/\phi_1, x_2\phi_2)$. Binned estimates from the anistropic model show good agreement with the line representing the model-based powered exponential estimate. The pairwise semivariance estimates, however, show a reasonable amount of deviation from the line. 

\begin{figure}[h!]
\begin{center}
\includegraphics[width=.49\textwidth, page=4]{semivar.pdf} \includegraphics[width=.49\textwidth]{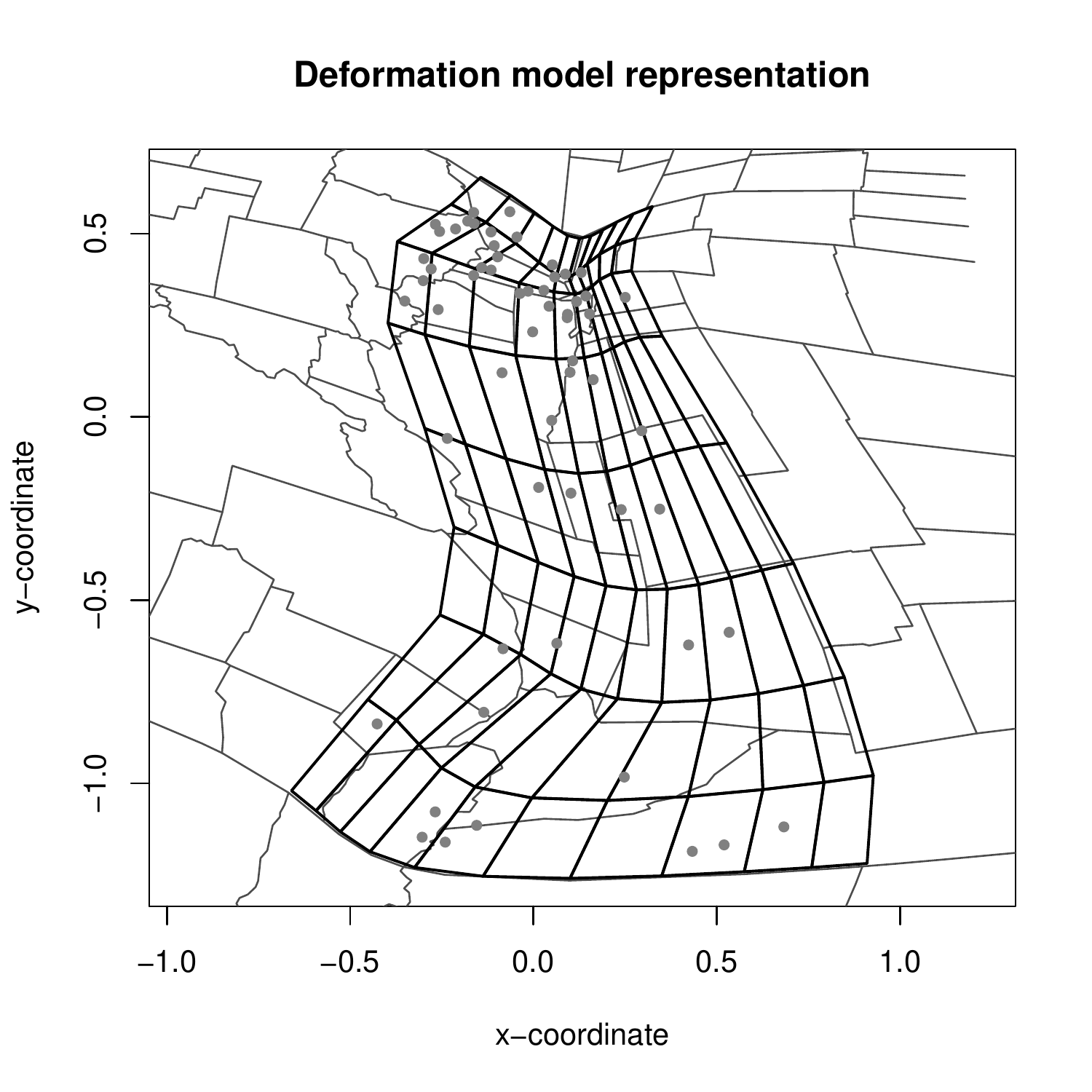} 
\end{center}
\caption{\label{F:rain:deform}Binned semivariance based on conventional anisotropic and deformed geostatistical models.}
\end{figure}

The spatial deformation model is used to allow for nonstationary covariance. This is specified so that $g_1$ and $g_2$ from \S\ref{S:meth:nonstat} are rank-12 thin plate regression splines. The resulting variogram for the model is shown in Figure \ref{F:rain:deform}. Marginal improvement can be seen over the anisotropic model as root mean square error (RMSE), which is defined with distance based on transformed coordinates and for pairwise estimates of $\rho_{ij}$ relative to their model-based counterparts, is slightly reduced. More compelling is the reduction in AIC, defined as in \citet[\S6.11.2]{wood-book}, which reduces from 2729385 for the anisotropic model to 2722578 for the deformation model. The resulting deformation is depicted in Figure \ref{F:rain:deform}, the most prominent feature of which is a decrease in grid cell areas from the southwest of the domain to the northeast, which corresponds to extreme rainfall events typically covering a larger area in the northeast than the southwest. Note that on this occasion a bijective mapping from $G$- to $D$-space arises without any penalty placed on folding.
 
\subsection{Extreme rainfall simulations}

\begin{figure}[h!]
\begin{center}
\includegraphics[width=.99\textwidth, page=1]{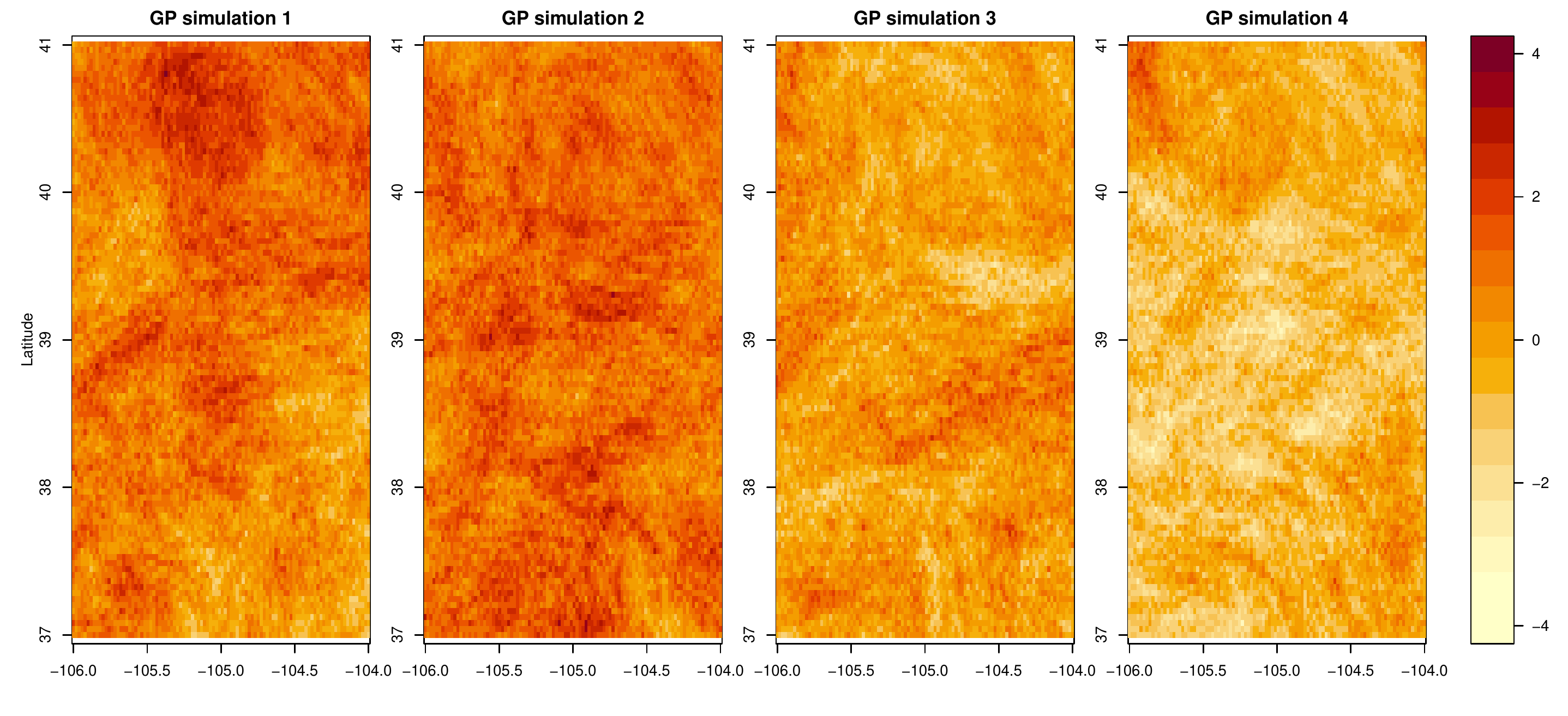}
\includegraphics[width=.99\textwidth, page=3]{sims.pdf}
\end{center}
\caption{\label{F:rain:sim1}Simulations for four arbitrary days on unit Gaussian scale (row 1) and transformed to original rainfall scale (row 2). For a given column, row 2 is derived from row 1. Grid cells in gray indicate simulated non-exceedances of the threshold from Figure \ref{F:rain:pars}.}
\end{figure}

Particularly useful for risk estimation is the ability to simulate extreme weather events. This process, for example, is often used in hazard modules of catastrophe models; see, e.g., \cite{grossi2005}. Figure \ref{F:rain:sim1} shows simulations of daily rainfall for four arbitrary days. Each day is represented by its original Gaussian process simulation and then its resulting rainfall simulation, where the latter is obtained by inverting the probability integral transformation of \eqref{E:pit}. Note that rainfall values are only generated when the threshold is exceeded; otherwise none is given. This is because non-exceedances of the threshold are treated as censored during model estimation. As a result, in two the simulated days, no rainfall values are simulated.
 
\begin{figure}[h!]
\begin{center}
\includegraphics[width=.99\textwidth]{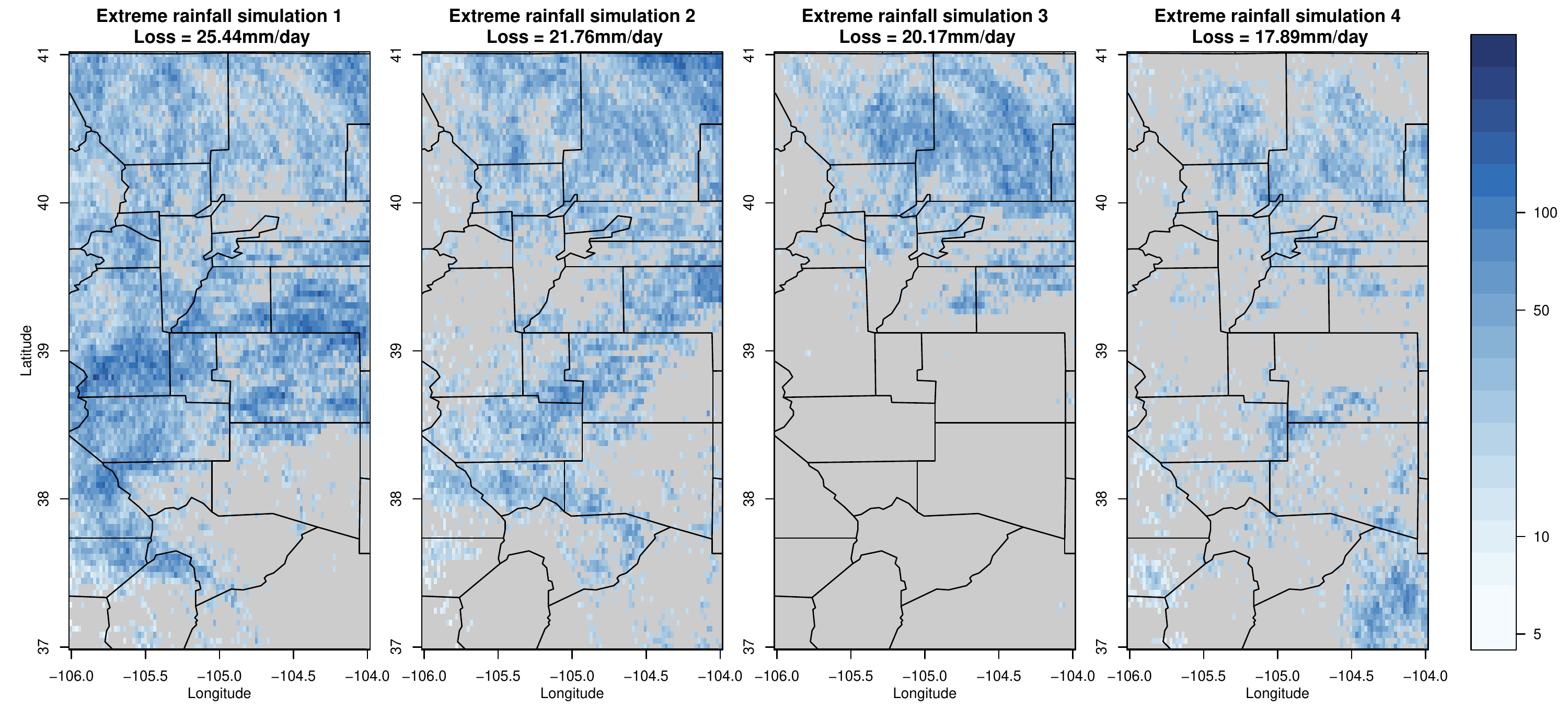}
\end{center}
\caption{\label{F:rain:sim2}Simulations for four days (between 1st April and 31st October) with largest total threshold excess over domain.}
\end{figure}

Rather more relevant to risk estimation is to consider the most extreme rainfall events. Therefore, 1000 years' events (for 1st April to 31st October) are simulated and a `loss' measure used to quantify their extremity. Here loss is defined as the mean rainfall excess (of the estimated threshold) per square kilometer. Figure \ref{F:rain:sim2} shows the four most extreme events and their losses. The event with largest loss has its highest rainfall values between 38 and 40 degrees latitude; the second largest loss seems to affect similar locations, but has less extreme rainfall between 38 and 40 degrees latitude; the third largest comprises an event over the northern half of the region; and the fourth largest comprises events in the southern, central and northern parts of the domain.

\section{Discussion} \label{discuss}

This work has developed an intuitive framework for representing nonstationary dependence for which objective inference is possible. Nonstationary dependence is achieved through spatial deformation, proposed by \cite{sampson1992}, or dimension expansion, developed in \cite{schmidt2011} and \cite{bornn2012}. The framework relies on splines and finite-rank representations of functions within the GAM setting so that results for such models, in particular automatic smoothing parameter estimation, allow for an essentially off-the-shelf approach to inference. In particular, this work allows tractable modelling for many locations using finite-rank deformation functions represented by thin plate regression splines, which additionally incorporate the constraints of \cite{smith1996} to avoid rotational invariance. This work also proposes a relatively simple yet intuitive numerical approach for avoiding non-bijective deformations, i.e., deformations in which $D$-space contains at least one fold, that can be applied to any deformation specification.

Although this work has presented methodology for nonstationary covariance by considering spatial processes, this is merely a special case within the GAM setting in which longitude and latitude are covariates. For example, spatial covariance can be allowed to vary with time through spatial deformations, or dimensions in the dimension expansion case, that vary with time. These can be achieved with three-dimensional time-varying $\bg$, formed, for example, through a tensor product of a two-dimensional function (as in \S\ref{S:meth:nonstat}) and a time-varying spline.

In \S\ref{S:solar:dimexp} the dimension expansion approach to inducing nonstationary covariance was explored through adding one and two extra dimensions. By relying on the GAM approach to representing these dimensions, each can be considered as ``smooths''. Therefore formal tests for deciding whether or not to retain smooths are applicable by, for example, considering $p$-values of smooths. Such testing is presented in detail in \citet[\S6.12]{wood-book}.

While the REML approach to inference brings objectivity to smoothing parameter estimation, if penalties are imposed to avoid folding in deformations some subjectivity is required for the parameters in the penalty functions in \S\ref{S:inf:fold}. In practice, the choice of $\delta$ has little influence on resulting estimated, provided it is large. The parameter $\epsilon$ has more effect. Since its resulting penalties depend on the chosen triangular tiling, of $G$-space, its value is simpler to specify relative to the area of the tiling's triangles or, as in \S\ref{solar-pen}, relative to the triangles' area once a conventional anisotropic model has been fitted. Choosing $\epsilon$ as small as possible to avoid folds in $D$-space then seems relatively robust, which can be judged from visualizations of grids at a appropriate resolution.

Although this work has been presented in the context of spatial modelling, it readily extends to statistical emulation, in which the relationship between a computer model's output and its inputs is represented by a statistical emulator. Often the emulator is a GP. Now let $\bx = (x_1, \ldots, x_p)$ denote an input to the emulator. A nonstationary covariance stucture could be allowed by assuming that $v(\bg(\bx), \cdot),$ where $v(\bg(\bx), \bg(\bx')) = \exp\{\minus\sum_{i=1}^p [g_i(x_i) - g_i(x_i')]^2\}$, assuming a Gaussian covariance structure. Note that this proposes each input to be deformed through a one-dimensional function, as opposed to spatial deformations in which each dimension is deformed through a two-dimensional function. Analogously to bijectivity in spatial deformations, monotonicity of each $g_d$ may want to be assumed. Approaches to achieving this are given in \cite{pya2015} and \citet[\S5.3.6]{wood-book}.

The application to Colorado rainfall forms the covariance matrix $\bf V$ from pairwise covariance estimates based on the bivariate Gaussian tail model of \cite{bor-ct}. The rainfall data contained relatively few missing values. However, if stations had large variation in their numbers of missing values, it would be inappropriate to use likelihood \eqref{gplik} without modification due to differing $T$. The GP's full likelihood could be used in this case, considering the Mahalanobis distance at each time point, but this would be computationally intensive. A compromise might be to group stations with similar numbers of non-missing values. Specifically, different groups could be formed of stations whose non-missing value count exceeds a given threshold, and $\bf V$ calculated for each. This would allow \eqref{gplik} to be partitioned according to an increasing sequence specified for $T$.

%

%
%
%
%

%

\bibliographystyle{chicago}

\bibliography{deform}
\end{document}